\documentclass[acmsmall,10pt]{acmart}
\usepackage[utf8]{inputenc}
\usepackage{amsmath}
\usepackage{amsfonts}
\usepackage{amsthm}
\usepackage{graphicx}

\usepackage{mathpartir}
\usepackage{listings}
\usepackage[british]{babel}

\usepackage{tikz}

\usepackage{stmaryrd}

\usepackage[autostyle]{csquotes}

\usepackage{todonotes}

\usepackage[all,british]{foreign}

\usepackage{booktabs}   %% For formal tables:
\usepackage{subcaption} %% For complex figures with subfigures/subcaptions
\newcommand{\codesize}{\fontsize{8}{7}\selectfont}

\newlength{\codeindent}
\lstdefinelanguage{sac}{morekeywords={with,fold,bool,}}

\lstdefinelanguage{FuncLang}
{
    keywords={Int,Float,Double,Bool,%
              imap, reduce, if, then, else, letrec, in,%
              filter},
    sensitive=true,
    showstringspaces=false,
    morecomment=[l]{;},
}
\lstdefinestyle{nonumbers}{numbers=none}

\newcommand{\map}{\textit{map}}
\newcommand{\imap}{\textit{imap}}
\newcommand{\reduce}{\textit{reduce}}
\newcommand{\filter}{\textit{filter}}
\newcommand{\lo}{\textit{islim}}
\newcommand{\evto}{\Downarrow}

\newcommand{\sac}{\textsc{SaC}}

\newcommand{\funclom}{\ensuremath{\lambda_\omega}}
\newcommand{\funclal}{\ensuremath{\lambda_\alpha}}

\newcommand{\cond}[3]{\textit{if}\ #1\ \textit{then}\ #2\ \textit{else}\ #3}
\newcommand{\letrec}[2]{\textit{letrec}\ #1\ \textit{in}\ #2}

\newcommand{\Gen}{\textit{Gen}}
\newcommand{\band}{\wedge}

\newcommand{\closure}[1]{\left\llbracket{} #1 \right\rrbracket}
\newcommand{\rulename}[1]{\text{\sc #1}}

\newcommand{\sep}{;\:}

\newcommand{\tuple}[1]{\langle#1\rangle}

\newcommand{\progvar}[1]{\textit{#1}}
\newcommand{\semvar}[2]{\ensuremath{#1_\text{#2}}}

\newcommand{\Exp}[1]{\text{\textbf{E}} (#1)}
\newcommand{\true}{\textit{true}}
\newcommand{\false}{\textit{false}}

\providecommand{\shp}[1]{\ensuremath{\big\lvert#1\big\rvert}}

\newcommand{\plusplus}{\mathrel{{+}+}}

\newenvironment{syntaxbox}[3][c]
{\begin{minipage}[#1]{#2}%\centering
#3
\vspace{3pt}
\hrule height1pt
\centerline{\vrule width1pt height 3pt\hfill \vrule width1pt height 3pt}
\begin{minipage}{\dimexpr\textwidth-4pt-1em}}
{\end{minipage}
\centerline{\vrule width1pt height 3pt\hfill \vrule width1pt height 3pt}
\hrule height1pt
\end{minipage}
\vspace{-3pt}}

\makeatletter
\newcommand{\labitem}[2]{%
\def\@itemlabel{\textbf{#1}}
\item
\def\@currentlabel{#1}\label{#2}}
\makeatother

\makeatletter\if@ACM@journal\makeatother
\acmJournal{PACMPL}
\acmVolume{1}
\acmNumber{1}
\acmArticle{1}
\acmYear{2017}
\acmMonth{1}
\acmDOI{10.1145/nnnnnnn.nnnnnnn}
\startPage{1}
\else\makeatother
\acmConference[ICFP'17]{ACM SIGPLAN International Conference on Functional Programming}%
{September 03--09, 2017}{Oxford, United Kingdom}
\acmYear{2017}
\acmISBN{978-x-xxxx-xxxx-x/YY/MM}
\acmDOI{10.1145/nnnnnnn.nnnnnnn}
\startPage{1}
\fi

\setcopyright{none}             %% For review submission
\begin{document}

%% Title information
\title[Transfinite Arrays]{A Lambda Calculus for Transfinite Arrays}
                                        %% [Short Title] is optional;
                                        %% when present, will be used in
                                        %% header instead of Full Title.
%\titlenote{with title note}            %% \titlenote is optional;
                                        %% can be repeated if necessary;
                                        %% contents suppressed with 'anonymous'
\subtitle{Unifying Arrays and Streams}  %% \subtitle is optional
%\subtitlenote{with subtitle note}       %% \subtitlenote is optional;
                                        %% can be repeated if necessary;
                                        %% contents suppressed with 'anonymous'

%% Author information
%% Contents and number of authors suppressed with 'anonymous'.
%% Each author should be introduced by \author, followed by
%% \authornote (optional), \orcid (optional), \affiliation, and
%% \email.
%% An author may have multiple affiliations and/or emails; repeat the
%% appropriate command.
%% Many elements are not rendered, but should be provided for metadata
%% extraction tools.

%% Author with single affiliation.
\author{Artjoms Šinkarovs}
%\authornote{with author1 note}          %% \authornote is optional;
                                        %% can be repeated if necessary
%\orcid{nnnn-nnnn-nnnn-nnnn}             %% \orcid is optional
\affiliation{
  \position{Research Associate}
  \department{School of Mathematical and Computer Sciences} %% \department is recommended
  \institution{Heriot-Watt University}            %% \institution is required
  \streetaddress{Heriot-Watt University}
  \city{Edinburgh}
  \state{Scotland}
  \postcode{EH14 4AS}
  \country{UK}
}
\email{first1.last1@inst1.edu}          %% \email is recommended

%% Author with two affiliations and emails.
\author{Sven-Bodo Scholz}
%\authornote{with author2 note}          %% \authornote is optional;
                                        %% can be repeated if necessary
%\orcid{nnnn-nnnn-nnnn-nnnn}             %% \orcid is optional
\affiliation{
  \position{Professor}
  \department{School of Mathematical and Computer Sciences} %% \department is recommended
  \institution{Heriot-Watt University}            %% \institution is required
  \streetaddress{Heriot-Watt University}
  \city{Edinburgh}
  \state{Scotland}
  \postcode{EH14 4AS}
  \country{UK}
}
\email{s.scholz@hw.ac.uk}         %% \email is recommended

%% Paper note
%% The \thanks command may be used to create a "paper note" ---
%% similar to a title note or an author note, but not explicitly
%% associated with a particular element.  It will appear immediately
%% above the permission/copyright statement.
%\thanks{with paper note}                %% \thanks is optional
                                        %% can be repeated if necessary
                                        %% contents suppressed with 'anonymous'

%% Abstract
%% Note: \begin{abstract}...\end{abstract} environment must come
%% before \maketitle command
\begin{abstract}
Array programming languages allow for concise and generic formulations of numerical
algorithms, thereby providing a huge potential for program optimisation such as
fusion, parallelisation, \etc{}  One of the restrictions that these
languages typically have is that the number of elements in every array
has to be finite.  This means that implementing streaming algorithms in
such languages requires new types of data structures, with 
operations that are not immediately compatible with existing
array operations or compiler optimisations.

In this paper, we propose a design for a functional language that natively
supports infinite arrays. We use ordinal numbers to introduce the notion of 
infinity in shapes and indices.  By doing so, we obtain a calculus
that naturally extends existing array calculi and, at the same time, allows
for recursive specifications as they are found in stream- and list-based settings.
Furthermore, the main language
construct that can be thought of as an $n$-fold \emph{cons} operator gives
rise to expressing transfinite recursion in data, something that lists
or streams usually do not support. This makes it possible to treat the proposed
calculus as a unifying theory of arrays, lists and streams.  We give
an operational semantics of the proposed language, discuss design choices
that we have made, and demonstrate its expressibility with several examples.
We also demonstrate that the proposed formalism preserves a number of well-known
universal equalities from array/list/stream theories, and discuss
implementation-related challenges.
\end{abstract}

%% 2012 ACM Computing Classification System (CSS) concepts
%% Generate at 'http://dl.acm.org/ccs/ccs.cfm'.
\begin{CCSXML}
<ccs2012>
<concept>
<concept_id>10003752.10010124.10010131.10010134</concept_id>
<concept_desc>Theory of computation~Operational semantics</concept_desc>
<concept_significance>500</concept_significance>
</concept>
</ccs2012>
\end{CCSXML}

\ccsdesc[500]{Theory of computation~Operational semantics}
%% End of generated code

\newcommand{\fixme}[1]{{ \color{red} \bf FIXME: #1}}

%% Keywords
%% comma separated list
\keywords{ordinals, arrays, semantics, functional languages}  %% \keywords is optional

%% \maketitle
%% Note: \maketitle command must come after title commands, author
%% commands, abstract environment, Computing Classification System
%% environment and commands, and keywords command.
\maketitle

%\tableofcontents{}
%\section*{Questions Thoughts and Facts}
%This is a meta-section that should be removed when all the questions will be
%resolved.
%
%Here are Thoughts:
%\begin{enumerate}
%    \item If we want to define recursive functions for all ordinals, we have to
%        support transfinite recursion properly.  To do this we need a built-in
%        predicate ``limit-ordinal''.
%    \item Presburger arithmetics with ordinals --- should be still decidable?
%    \item Co-induction can be often replaced with transfinite induction.
%\end{enumerate}

\section{Introduction}

Array-based computation offers many appealing properties when dealing with
large amounts of homogeneous data. All data can be accessed in $O(1)$ time, storage
is compact, and array programs typically lend themselves
to data-parallel execution. Another benefit arises from the fact that many
applications naturally deal with data that is structured along several independent
axes of linearly ordered elements.

Besides these immediate benefits, the structuring facilities of arrays
offer significant opportunities for developing elaborate array calculi, such
as Mullin's $\psi$-calculus~\cite{LMRMullin:moa,Mullin94},
Nial~\cite{Jenkins88} and the many
APL-inspired array languages~\cite{ibm:aplxm6,RBernecky:nirf,
RKWHui:jdictionary}.
These calculi benefit programmers in several ways.

Firstly, they improve \emph{programmer productivity}.
Array calculi provide the grounds for a rich set of generic operators.
Having these readily available, programmers can compose programs more quickly;
the resulting algorithms are typically more versatile than implementations
that manipulate arrays directly on an element-by-element basis.

Secondly, array calculi help improving \emph{program correctness}.
The operators that build the foundation of any given array calculus come with a wealth
of properties that manifest in equalities that hold for all arrays.
With these properties available, programming becomes less error-prone,
\eg{} avoidance of out-of-bound accesses through the use of array-oriented operators.
It also becomes easier to reason formally about programs and their behaviour.

Finally, array calculi help compilers to \emph{optimise and parallelise} programs.
The aforementioned equalities can also be leveraged when it comes to high-performance
parallel executions.
They allow compilers to restructure both algorithms and data structures,
enabling improvements such as better data
locality~\cite{Bondhugula:2008:PAP,Grelck2000},
better vectorisation~\cite{Allen:1987:ATF,poly-autovec,ccpe-data-layouts}, and streaming through
accelerator devices~\cite{Alias:2013:ORA,Guo:2011:BGP}.

These advantages have inspired many languages and their attendant tool chains
including various APL
implementations~\cite{RBernecky:nirf,Dyalog:ReferenceManual,%
ibmapl2:langref,RKWHui:jdictionary},
parallel arrays in Haskell~\cite{Keller2010},
push-pull arrays in Haskell~\cite{pushpull}, and arrays in
\sac{}~\cite{GerlckSchoIJPP06}
and Futhark~\cite{Henriksen:2017:FPF:3062341.3062354}.

Despite generic specification and the ability to stream
finite arrays, array languages typically cannot deal with infinite streams.
If an existing application for finite arrays needs to be extended
to deal with infinite streams, a complete code rewrite is often required,
even if the algorithmic pattern applied to the array elements is unchanged.
Retrofitting such streaming often obfuscates the core algorithm, and adds
overhead when the algorithm is applied to finite streams.
If the overhead is non-negligible, both code versions need to be
maintained and, if the finiteness of the data
is not known \apriori{}, a dynamic switch between them is required.

Streaming style leads to a completely different way of thinking
about data. Very similar to programming on lists, traditional
streaming deals with
individual recursive acts of creation or consumption. This is
appealing, as elegant recursive data definitions become
possible and element insertion and deletion can be implemented
efficiently, which is difficult to achieve in a traditional array setting.
Another aspect of streams is the inevitable temporal existence
of parts of streams or lists, which quite well matches
the lazy regime prevalent in list-based languages,
but is usually at odds with obtaining high-performance
parallel array processing.

In this paper, we try to tackle this limitation of array languages when it comes to infinite
structures. Specifically, we look at extending array languages in a consistent way, to support
streaming through infinite dimensions.
Our aim is to avoid switching to a traditional
streaming approach, and to stay within the array paradigm,
thereby making it possible to use the same algorithm specification
for both finite and infinite inputs,
possibly maintaining the benefits of a given underlying array calculus.
We also hope that excellent parallel performance can be maintained
for the finite cases, and that typical array-based optimisations can be
applied to both the finite and infinite cases.

We start from an applied $\lambda$-calculus supporting
finite $n$-dimensional arrays, and investigate extensions to
support infinite arrays.
The design of this calculus aims to provide a solid basis
for several array calculi, and to facilitate compilation to high-performance parallel code.

In extending the calculus to deal with infinities, we pay particular
attention to the algebraic properties that are present in the finite
case, and how they translate into the infinite scenario.  The key insight here
is that when ordinals are used to describe shapes and indices of arrays,
many useful properties can be preserved.
Borrowing from the nomenclature of ordinals, we refer to these ordinal-indexed
infinite arrays as \emph{transfinite arrays}.
We identify and minimize requisite semantic extensions and modifications.
We also look into the relationship between the resulting array-based $\lambda$-calculus
and classical streaming.
Finally, we look at several examples, and discuss implementation issues.

The individual contributions of this paper are as follows:
\begin{enumerate}
    \item We define an applied $\lambda$-calculus on finite arrays, and its
          operational semantics. The calculus is a rather generic core language that
          implicitly supports several array calculi as well as compilation
          to highly efficient parallel code.

    \item We expand the $\lambda$-calculus to support infinite arrays and show
          that the use of ordinals as indices and shapes creates a wide range of universal
          equalities that apply to finite and transfinite arrays alike.

    \item We show that the proposed calculus also maintains many streaming properties
          even in the context of transfinite streaming.

    \item We show that the proposed calculus inherently supports transfinite
          recursion. Several examples are contrasted to traditional list-based
          solutions.

    \item We describe a prototypical implementation\footnote{
        The implementation is freely available at
        \url{https://github.com/ashinkarov/heh}.
        %The implementation is provided in the anonymous supplementary materials.
        }, and briefly discuss
        the opportunities and challenges involved.
\end{enumerate}

We start with a description of the finite array calculus and naive extensions for infinite
arrays in Section~2, before presenting the ordinal-based approach and its potential
in Sections~3--5.
Section~6 presents our prototypical implementation.
Related work is discussed in Section~7; we conclude in Section~8.

\section{Extending Arrays to Infinity}

We define an idealised, data-parallel array language, based on
an applied $\lambda$-calculus that we call \funclal{}.
The key aspect of \funclal{} is built-in support for shape- and rank-polymorphic
array operations, similar to what is available in APL~\cite{APL}, J~\cite{J},
or \sac{}~\cite{GerlckSchoIJPP06}.

In the array programming community, it is well-known~\cite{Jenkins1991,jenkins99}
that basic design choices made in a language have an impact on the array
algebras to which the language adheres. While we believe that our proposed
approach is applicable within various array algebras, we chose one
concrete setting for the context of this paper.  We follow the \textit{design
decisions} of the functional array language \sac{}, which are
compatible with many array languages, and which were taken directly from
K.E. Iverson's design of APL\@.

\begin{description}
    \labitem{DD~1}{design:array} \textit{All expressions in \funclal{} are
        arrays.} Each array has a shape which defines how components
        within arrays can be selected.
    \labitem{DD~2}{design:scalar} \textit{Scalar expressions, such as constants
        or functions, are 0-dimensional objects with empty shape.} Note that
        this maintains the property that all arrays consist of as many elements as
        the product of their shape, since the product of an empty shape is
        defined through the neutral element of multiplication, \ie{} the number 1.
    \labitem{DD~3}{design:rect} \textit{Arrays are rectangular --- the index
        space of every array forms a hyper-rectangle.} This allows the shape of
        an array to be defined by a single vector containing the element
        count for each axis of the given array.
    \labitem{DD~4}{design:nested} \textit{Nested arrays that cater for
        inhomogeneous nesting are not supported.  Homogeneously nested array
        expressions are considered isomorphic with non-nested higher-dimensional
        arrays.} Inhomogeneous nesting, in principle, can be supported by adding
        dual constructs for enclosing and disclosing an entire array into a singleton,
        and vice versa. {}\ref{design:scalar} implies that functions
        and function application can be used for this purpose.
    \labitem{DD~5}{design:empty} \textit{\funclal{} supports infinitely many
        distinct empty arrays that differ only in their shapes.} In the
        definition of array calculi, the choice whether there is only one empty array
        or several has consequences on the universal equalities that hold.  While a
        single empty array benefits value-focussed equalities, structural equalities
        require knowledge of array shapes, even when  those arrays are empty.
        In this work, we assume an
        infinite number of empty arrays; any array with at least one shape element
        being $0$ is empty.  Empty arrays with different shape are considered
        distinct.  For example, the empty arrays of shape $[3,0]$ and
        $[0]$ are different arrays.
\end{description}

\subsection[Syntax Definition and Informal Semantics of lambda-alpha]%
{Syntax Definition and Informal Semantics of \funclal{}}
\begin{figure}[h!]
%\begin{syntaxbox}[b]{.7\textwidth}{}
\begin{tabular}{@{}ll@{}}
\begin{minipage}[t]{.49\textwidth}
\[
\begin{array}{ccll}
    c &::=& 0,1,\dots, & \text{(numbers)} \\
      &|& \true{}, \false{} & \text{(booleans)} \\
    \\
    e &::=& c & \text{(constants)}\\
      &|&   x &  \text{(variables)}\\
      &|&   \lambda x.e & \text{(abstractions)}\\
      &|&   e\ e & \text{(applications)}\\
      &|&   \cond{e}{e}{e} & \text{(conditionals)}\\
      &|&   \letrec{x = e}{e} & \text{(recursive let)}\\
      &|&   e+e, \dots & \text{(built-in binary)} \\
      &|&   [e, \dots, e] & \text{(array constructor)}\\
      &|&   e . e & \text{(selections)}\\
      &|&   |e| & \text{(shape operation)} \\
      &\sim
\end{array}
\]
\end{minipage}
&
\begin{minipage}[t]{.49\textwidth}
\[
\begin{array}{ccll}
     &\sim \\
      &|&   \reduce\ e\ e\ e & \text{(reduction)} \\
      &|&   \imap\ s
            \begin{cases}
                g_1:& e_1, \\
                \qquad\dots&\\
                g_n:& e_n
            \end{cases}& \text{(index map)}\\
      \\
    s &::=& e & \text{(scalar imap)} \\
      &|& e|e & \text{(generic imap)}\\
    g &::=& (e\ \texttt{<=}\ x\ \texttt{<}\ e) & \text{(index set)} \\
      &|&  \_(x) & \text{(full index set)}
\end{array}
\]
\end{minipage}
\end{tabular}
%\end{syntaxbox}
\caption{\label{fig:syntax}The syntax of \funclal}
\end{figure}

We define the syntax of \funclal{} in Fig.~\ref{fig:syntax}.
Its core is an untyped, applied $\lambda$-calculus.
Besides scalar constants, variables, abstractions and applications,
we introduce conditionals, a recursive let operator and some basic
functions on the constants, including
arithmetic operations such as \texttt{+}, \texttt{-}, \texttt{*}, \texttt{/},
a remainder operation denoted as \texttt{\%},
and comparisons \texttt{<}, \texttt{<=}, \texttt{=}, \etc{}
The actual support for arrays as envisioned by the aforementioned design principles
is provided through five further constructs: array construction, selection,
shape operation, \reduce{} and \imap{} combinators.

All arrays in \funclal{} are immutable.  Arrays can be constructed by using
potentially nested sequences of scalars in square
brackets. For example, $[1, 2, 3, 4 ]$ denotes a four-element vector, while
$[[1, 2], [3, 4]]$ denotes a two-by-two-element matrix.
We require any such nesting to be homogeneous, for adherence to~\ref{design:nested}.
For example, the term $[[1, 2], [3]]$
is irreducible, so does not constitute a value.

The dual of array construction is a built-in operation for element selection, denoted
by a dot symbol, used as an infix binary operator between an array to select from,
and a valid index into that array.
A valid index is a vector containing as many elements as the array has
dimensions; otherwise it is undefined.
\begin{mathpar}[]
 [1, 2, 3, 4].[0] = 1 \and
 [[1, 2], [3, 4]].[1, 1] = 4 \and
 [[1, 2], [3, 4]].[1] = \bot
\end{mathpar}

The third array-specific addition to \funclal{} is the primitive \emph{shape} operation,
denoted by enclosing vertical bars. It is applicable to arbitrary expressions,
as demanded by~\ref{design:array}, and it returns the shape of its argument as a
vector, leveraging~\ref{design:rect}.  For our running examples, we obtain:
\(
    \shp{[1, 2, 3, 4]} = [4]\ \text{and}\
    \shp{[[1, 2], [3, 4]]} = [2, 2]
\).
{}\ref{design:empty} and {}\ref{design:scalar} imply that we have:
\begin{mathpar}
    \shp{[]} = [0] \and
    \shp{[[]]} = [1, 0] \and
    \shp{\true{}\,} = [] \and
    \shp{42} = [] \and
    \shp{\lambda x.x} = []
\end{mathpar}

\funclal{}  includes a \reduce{} combinator which in essence,
it is a variant of \emph{foldl}, extended to allow for
multi-dimensional arrays instead of lists.
\reduce{} takes three arguments: the binary function, the neutral element
and the array to reduce.
For example, we have:
\begin{mathpar}
   \reduce\ (+)\ 0\ [[1,2],[3,4]] = ((((0 + 1) + 2) + 3) + 4)
\end{mathpar}
assuming row-major traversal order.  This allows for shape-polymorphic
reductions such as:
\begin{lstlisting}[mathescape, style=nonumbers]
sum $\equiv$ $\lambda$a.reduce ($\lambda$x.$\lambda$y.x+y) 0 a    ; also works for scalars and empty arrays
\end{lstlisting}

The final, and most elaborate, language construct is the \imap{} (index map) construct.
It bears some similarity to the classical map operation,
but instead of mapping a function over the
elements of an array, it constructs an array by mapping a function over all legal indices into
the index space denoted by a given shape expression\footnote{
    For readers familiar with Haskell: the \imap{} defined here derives
the index space from a shape expression. It does
not require an argument array of that shape.}.
Added flexibility is obtained by supporting a piecewise definition of
the function to be mapped.
Syntactically, the \imap{}-construct starts out with the keyword \texttt{imap},
followed by a description of the result shape (rule $s$ in Fig.~\ref{fig:syntax}).
The shape description is followed by a curly bracket that precedes the definition
of the mapping function.
This function can be defined piecewise by providing a set of
index-range expression pairs.
We demand that the set of index ranges constitutes a partitioning of the overall
index space defined through the result shape expression, \ie{} their union covers
the entire index space and the index ranges are mutually disjoint.
We refer to such index ranges as \emph{generators} (rule $g$ in Fig.~\ref{fig:syntax}),
and we call a pair of a generator and its subsequent expression a \emph{partition}.
Each generator defines an index set and a variable (denoted by $x$ in rule $g$
in Fig.~\ref{fig:syntax}) which serves as the formal parameter of the function to
be mapped over the index set.  Generators can be defined in two ways: by means
of two expressions which must evaluate to vectors of the same shape, constituting
the lower and upper bounds of the index set, or by using the underscore notation which
is syntactic sugar for the following expansion rule:
\begin{lstlisting}[mathescape, style=nonumbers]
(imap s { _(iv) ...) $\equiv$  (imap s { [$\underbrace{0,...,0}_{n}$] <= iv < s: ...)
\end{lstlisting}
assuming that $\shp{s} = [n]$.  The variable name of a generator can be referred to in
the expression of the corresponding partition.

The \texttt{<=} and  \texttt{<} operators in the generators
can be seen as element-by-element array counterparts of the corresponding scalar operators
which, jointly, specify sets of constraints on the indices described by the generators.
As the index-bounds are vectors, we have:
\[v_1 \mathrel{\texttt{<=}} v_2
    \implies \shp{v_1}.[0] = \shp{v_2}.[0]
       \band \forall 0 \mathrel{\texttt{<=}} i < \shp{v_1}.[0]:
                                v_1.[i] \mathrel{\texttt{<=}} v_2.[i]
\]
In the rest of the paper, we use the same element-wise extensions
for scalar operators, denoting the non-scalar versions with
dot on top: $c = a \dot{+} b \implies c.i = a.i + b.i$.  This often
helps to simplify the notation\footnote{A formal definition of
    the extended operator is:
    $(\dot{\oplus}) \equiv
        \lambda a.\lambda b.
            \imap\ |a|\ \{\_(\progvar{iv}):
                          a.\progvar{iv} \oplus b.\progvar{iv}$
    where $\oplus \in \{+,-,\cdots\}$.}.

As an example of an \imap{}, consider an element-wise increment of an array $a$ of shape $[n]$.
While a classical \map{}-based definition can be expressed as $\map\ (\lambda x.x+1)\ a$,
using \imap{}, the same operation can be defined as:
\begin{lstlisting}[mathescape, style=nonumbers]
imap [n] { [0] <= iv < [n]: a.iv + 1
\end{lstlisting}

Having mapping functions from indices to values rather than values to values
adds to the flexibility of the construct.
Arrays can be constructed from shape expressions without requiring an array
of the same shape available:
\begin{lstlisting}[mathescape, style=nonumbers]
imap [3,3] { [0,0] <= iv < [3,3]: iv.[0]*3 + iv.[1]
\end{lstlisting}
defines a 2-dimensional array $[[0,1,2],[3,4,5],[6,7,8]]$.
Structural manipulations can be defined conveniently as well.
Consider a \emph{reverse} function, defined as follows:
\begin{lstlisting}[mathescape, style=nonumbers]
reverse $\equiv$ $\lambda$a.imap |a| { [0] <= iv < |a|: a.(|a|$\dot{-}$iv$\dot{-}$[1])
\end{lstlisting}
In order to express this with \map{}, one needs to construct an intermediate
array, where indices of $a$ appear as values.  Note also that the explicit shape
of the \imap{} construct makes it possible to define shape-polymorphic
functions in a way similar to our definition of $reverse$.
An element-wise increment for arbitrarily shaped arrays can be defined as:
\begin{lstlisting}[mathescape, style=nonumbers]
increment $\equiv$ $\lambda$a.imap |a| { _(iv): a.iv + 1 ; also works for scalars & empty arrays
\end{lstlisting}

{}\ref{design:nested} allows \imap{} to be used for expressing operations in
terms of $n$-dimensional sub-structures.  All that is required for this is that the expressions on
the right hand side of all partitions evaluate to non-scalar values.  For
example, matrices can be constructed from vectors.  Consider the following
expression:
\begin{lstlisting}[mathescape, style=nonumbers]
imap [n] { [0] <= iv < [n]: [1,2,3,4] ; non-scalar partitions (incorrect attempt)
\end{lstlisting}
Its shape is $[n,4]$; however, this shape no longer can be computed without
knowing the shape of at least one element.  If the overall result array
is empty, its shape determination is a non-trivial problem.
To avoid this situation, we
require the programmer to specify the result shape by means of two shape
expressions separated by a vertical bar: see the rule \text{(generic imap)} in
Fig.~\ref{fig:syntax}.  We refer to these two shape expressions as the
\emph{frame shape} which specifies the overall index range of the \imap{}
construct as well as the \emph{cell shape} which defines the shape of all
expressions at any given index. The concatenation of those two shapes is the
overall shape of the resulting array.  For more discussions related to the
concepts of frame and cell shapes,
see~\cite{RBernecky:operencl,RBernecky:funrank,RBernecky:apljhiperf}.  The above
\imap{} expression therefore needs to be written as:
\begin{lstlisting}[mathescape, style=nonumbers]
imap [n]|[4] { [0] <= iv < [n]: [1,2,3,4] ; non-scalar partitions (correct)
\end{lstlisting}
to be a legitimate expression of \funclal{}.  The \text{(scalar imap)} case in
Fig.~\ref{fig:syntax}, which we use predominantly in the paper, can be seen as
syntactic sugar for the generic version, with the second expression being an
empty vector.

% \paragraph{Reduce}
% \funclal{}  includes a \reduce{}-construct for implementing reductions.
% In essence, it is a variant of \emph{foldl} extended to allow for
% multi-dimensional arrays instead of lists.
% For example, we have:
% \begin{mathpar}
%    \reduce\ (+)\ 0\ [[1,2],[3,4]] = ((((0 + 1) + 2) + 3) + 4)
% \end{mathpar}
% assuming row-major traversal order.  This allows for shape polymorphic
% reductions such as:
% \begin{lstlisting}[mathescape, style=nonumbers]
% sum $\equiv$ $\lambda$a.reduce (+) 0 a
% \end{lstlisting}
%

\subsection[Formal Semantics of lambda-alpha]%
{Formal Semantics of \funclal{}}

In this section, we offer a brief overview of the semantics. A complete semantics
can be found in~\cite{lamom-sem}.

In \funclal{}, evaluated arrays are pairs of shape and element
tuples. A shape tuple consists of numbers, and an element tuple consists
of numbers, booleans or functions closures.  We denote pairs
and tuples, as well as element selection and concatenation on them, using the
following notation:
\begin{mathpar}
    \vec{a} = \tuple{a_1, \dots, a_n}
    \implies
    \vec{a}_i = a_i
    \and
    \tuple{a_1, \dots, a_n} \plusplus \tuple{b_1, \dots, b_m}
    = \tuple{a_1, \dots, a_n, b_1, \dots, b_m}
\end{mathpar}
To denote the product of a tuple of numbers, we use the following notation:
\[
    \vec{s} = \tuple{s_1, \dots, s_n}
    \implies \otimes \vec{s} = s_n \cdot \cdots s_1 \cdot 1
\]
%Note that we multiply elements in the reverse order, which is important as
%ordinal multiplication is not commutative.  Also,
When a tuple is empty, its product is one.
An array is rectangular, so its shape vector specifies the extent of each axis.
The number of elements of each array is finite.
Element vectors contain all the elements in a linearised form. While the
reader can assume row-major order, formally, it suffices that a fixed
linearisation function $F_{\vec{s}}$ exists which,
given a shape vector $\vec{s} = \tuple{s_1, \dots, s_n}$, is a
bijection between indices $\{\tuple{0,\dots, 0},\dots,\tuple{s_1-1,\dots,s_n-1}\}$
and offsets of the element vector: $\{1, \dots, \otimes\vec{s}\}$.
Consider, as an example, the array $[[1,2],[3,4]]$, with $F$ being row-major order.
This array is evaluated into the shape-tuple element-tuple pair
\(
    \tuple{\tuple{2,2},\tuple{1,2,3,4}}
\).
Scalar constants are arrays with empty shapes. We have
$5$ evaluating to $\tuple{\tuple{},\tuple{5}}$.  The same holds for booleans
and function closures:
\true{} evaluates to $\tuple{\tuple{}, \tuple{\true{}}}$ and
$\lambda x.e$ evaluates to $\tuple{\tuple{}, \tuple{\closure{\lambda x.e, \rho}}}$.

$F$ is an invariant to the presented semantics.  In finite cases, the usual
choices of $F$ are row-major order or column-major order.  In infinite cases,
this might be not the best option, and one could consider space-filling curves
instead.
$F$ is only relevant for two operations; the creation of array values and
the selection of elements from it.
Selections relate the indices of the index vectors to the axes of the arrays
following the order of nesting and starting with the index 0 on each level.
We have:
\(
    [[1,2],[3,4]]\ [1,0] = 3
\),
Assuming $F$ is row-major, $F_{\tuple{2,2}}(\tuple{1,0})$ equals $2$
which, when used as index into $\tuple{\tuple{2,2},\tuple{1,2,3,4}}$ returns
the intended result $3$.

The inverse of $F$ is denoted as
$F^{-1}_{\vec{s}}$ and for every legal offset $\{1, \dots, \otimes\vec{s}\}$
it returns an index vector for that offset.

\paragraph{Deduction rules}
To define the operational semantics of \funclal{}, we use a \textit{natural
semantics}, similar to the one described in~\cite{Kahn:NaturalSemantics}.
To make sharing more visible, instead of a single environment $\rho$ that
maps names to values, we introduce a concept of storage;
environments map names to pointers and storage maps pointers to values.
Environments are denoted by $\rho$ and are ordered lists of name-pointer pairs.
Storage is denoted by $S$ and consists of
an ordered list of pointer-value pairs.

Formally, we construct storage and environments as lists of pointer-value and
variable-pointer bindings, respectively, using comma to denote extensions:
\[
    S ::= \emptyset\ |\ S, p \mapsto v\qquad
    \rho ::= \emptyset\ |\ \rho, x \mapsto p
\]
A look-up of a storage or an environment is performed \textit{right to left}
and is denoted as $S(p)$ and $\rho(x)$, respectively. Extensions are denoted
with comma.
%:
%\[
%    S' = S, p \mapsto v \qquad \rho' = \rho, x \mapsto p
%\]
%When extending a storage, we implicitly assume that the name of the pointer
%is always fresh.  When extending environments, we preserve the name coming
%from $\lambda$ abstraction or \emph{letrec}, but otherwise when we use a
%variable to temporarily assign a pointer, the variable name is assumed
%to be fresh.
Semantic judgements can take two forms:
\begin{mathpar}
    S;\rho \vdash e \evto S'\sep p\qquad
    S;\rho \vdash e \evto S'\sep p \Rightarrow v
\end{mathpar}
where $S$ and $\rho$ are initial storage and environment and $e$ is a
\funclal{} expression to be evaluated.  The result of this evaluation ends up
in the storage $S'$ and the pointer $p$ points to it.  The latter form of a
judgement is a shortcut for: \(S;\rho \vdash e \evto S'\sep p \band S'(p) = v\).

\paragraph{Values}
The values in this semantics are constants (including arrays) and $\lambda$-closures
which contain  the $\lambda$ term and the environment where this term shall be
evaluated:
\[
    \tuple{\tuple{\dots},\tuple{\dots}}\qquad
    \tuple{\tuple{},\tuple{\closure{\lambda x.e, \rho}}}
\]

\paragraph{Meta-operators}
Further in this section we use the following meta-operators:
\begin{description}
    \item[$\Exp{v}$] Lift the internal representation of a vector or a number
        into a valid \funclal{} expression.  For example: $\Exp{5} = 5$,
        $\Exp{\tuple{1,2,3}} = [1,2,3]$, \etc{}
    \item[$\tuple{\vec{s}, \_}$] We use underscore to omit the part of a data
        structure, when binding names.  For example:
        $S\sep p \Rightarrow \tuple{\vec{s}, \_}$ refers to binding the variable
        $\vec{s}$ to the shape of $S(p)$ which must be a constant.
%    \item{$\text{Nxt}(\vec{u})$} The next value after $\vec{u}$ in the lexicographical
%        ordering of indices.  The orderi
\end{description}

\subsection{Core Rules}

In \funclal{}, the rules for the $\lambda$-calculus core, \ie{} constants, variables,
abstractions and applications are straightforward adaptations of the standard rules
for strict functional languages to our notation with storage and pointers:

\begin{mathpar}
    \inferrule[Const-Scal]
        {
            c\ \textit{is scalar}
        }
        {
            S;\rho \vdash c \evto S_1,p\mapsto \tuple{\tuple{}, \tuple{c}}\sep p
        }
    \and
    \inferrule[Var]
        {
            x \in \rho \\
            \rho(x) \in S
        }
        {
            S;\rho \vdash x \evto S\sep \rho(x)
        }
    \and \\
    \inferrule[Abs]
        {
        }
        {
            S;\rho \vdash \lambda x.e \evto
            S,p\mapsto \tuple{\tuple{}, \tuple{\closure{\lambda x.e, \rho}}}\sep p
        }
    \and
    \inferrule[App]
        {
            S;\rho \vdash e_1 \evto S_1\sep p_1
            \Rightarrow \tuple{\tuple{},\closure{\lambda x.e, \rho_1}} \\\\
            S_1;\rho \vdash e_2 \evto S_2\sep p_2 \\
            S_2;\rho_1, x\mapsto p_2 \vdash e \evto S_3\sep p_3
        }
        {
            S;\rho \vdash e_1\ e_2 \evto S_3\sep p_3
        }
\end{mathpar}

As an illustration, consider the evaluation of $(\lambda x.x)\ 42$:
\[
\begin{array}{@{}lrl}
    \emptyset\sep\emptyset
    & (\lambda x. x)\ 42
    & \rulename{Abs} \\
    S_1 = p_1 \mapsto \tuple{\tuple{}, \closure{\lambda x.x, \emptyset}}\sep \emptyset
    & p_1\ 42
    & \rulename{Const-Scal} \\
    S_2 = S_1, p_2 \mapsto \tuple{\tuple{}, \tuple{42}}\sep \emptyset
    & p_1\ p_2
    & \rulename{App} \\
    S_2\sep x\mapsto p_2
    & x
    & \rulename{Var} \\
    S_2\sep\emptyset
    &p_2
    &\square
\end{array}
\]

We start with an empty storage and an empty environment.
The outer application demands that the \rulename{App}-rule be used. It
enforces three computations: the evaluation of the function, the evaluation
of the argument and the evaluation of the function body with an appropriately
expanded environment.
The function is evaluated by the \rulename{Abs}-rule which adds
a closure \(p_1\mapsto \tuple{\tuple{}, \closure{\lambda x.x, \emptyset}}\) to the storage and returns
the pointer $p_1$ to it.
The argument is evaluated by the \rulename{Const-Scal}-rule which adds
\(p_2 \mapsto \tuple{\tuple{},\tuple{42}}\) to the storage and returns $p_2$.
Finally, the \rulename{App}-rule demands the evaluation of the body of the function
with an environment \(\rho_1= x\mapsto p_2\).
The body being just the variable $x$, the \rulename{Var}-rule gives us $S_2\sep p_2$ as
final result.
% \begin{mathpar}
% S_2; \rho_1 \vdash x \evto S_2; p_2 \qquad \text{where }
% S_2 = \{p_1\mapsto \closure{\lambda x.x, \emptyset}, p_2 \mapsto \tuple{\tuple{},\tuple{42}}\}
% \text{ and } \rho_1= x\mapsto p_2.
% \end{mathpar}
%
% Putting these results together, the \textsc{App}-rule delivers the final result as:
% \begin{mathpar}
% S; \rho \vdash \lambda x.x \ [1,2,3] \evto S_2;p_2 \qquad \text{where}\
% S= \emptyset,\ \rho = \emptyset,\ \text{and}\  S_2 =
%   \{p_1 \mapsto \closure{\lambda x.x, \emptyset}, p_2 \mapsto \tuple{\tuple{},\tuple{42}}\}
% \end{mathpar}

The rules for array constructors and array selections are rather straightforward as well. Both
these constructs are strict:

\begin{mathpar}
    \inferrule[Imm-Array]
        {
            n \geq 1 \\
            \mathop{\forall}\limits_{i=1}^n S_i;\rho \vdash c_i \evto S_{i+1}\sep p_i \\
            P = \tuple{p_1, \dots, p_n} \\
            %\text{AllFiniteShape}(S_{n+1},P) \\
            \text{AllSameShape}(S_{n+1},P) \\
            S' = S_{n+1}, p_o \mapsto \tuple{\tuple{1},\tuple{n}},
                          p_i \mapsto S_{n+1}(p_1) \\
            S',\rho \vdash
            \imap_1\ p_o|p_i\ \left\{\tuple{i{-}1}
                                \mapsto p_i\  |\  i \in \{1, \dots, n\}\right\}
            \evto S''\sep p
        }
        {
            S_1;\rho \vdash [c_1, \dots, c_n] \evto S''\sep p
        }
    \and
    \inferrule[Imm-Array-empty]
        {
        }
        {
            S;\rho \vdash [] \evto S,p\mapsto \tuple{\tuple{0}, \tuple{}}\sep p
        }
    \and
    \inferrule[Sel-strict]
        {
            S;\rho \vdash i \evto S_1\sep p_i
            \Rightarrow \tuple{\tuple{d}, \vec{\imath}} \\
            S_1;\rho \vdash a \evto S_2\sep p_a
            \Rightarrow \tuple{\vec{s}, \vec{a}} \\
            k = F_{\vec{s}}(\vec{\imath})
        }
        {
            S;\rho \vdash a.i
            \evto S_3, p\mapsto \tuple{\tuple{},\tuple{\vec{a}_k}}\sep p
        }
\end{mathpar}

Empty arrays are put into the storage with shape $[0]$ (\textsc{Imm-Array-empty}-rule).
Non-empty arrays (\textsc{Imm-Array}-rule) evaluate all the components and ensure
that they are all of the same finite shape.  Subsequently,
we assemble evaluated components into the resulting array value ensuring that the flattening
adheres to $F$. This is achieved by using an auxiliary term $\imap_1$.
It takes the form $\imap_1\ p_o|p_i\ \{\vec\imath^{\,1} \mapsto p_{\vec\imath^{\,1}},
                               \dots,
                               \vec\imath^{\,n} \mapsto p_{\vec\imath^{\,n}}\}$
where $p_o$ and $p_i$ are pointers to frame and cell shapes, and the set
$\{\vec\imath^{\,1} \mapsto p_{\vec\imath^{\,1}},
                               \dots,
                               \vec\imath^{\,n} \mapsto p_{\vec\imath^{\,n}}\}$
contains pairs of frame-shape indices and value pointers for all legal
indices into the frame shape.
The formal definition of the deduction rule for $\imap_1$ is provided
in~\cite[Sec~2.1.1]{lamom-sem}.

The rule for selection (\textsc{Sel-strict}-rule) first evaluates
the array we are selecting from, and the index vector specifying
the array index we wish to select.  Then, we compute the offset into the data vector
by applying $F$ to the index vector.  Finally, we get the scalar value at
the corresponding index.  When applying $F$, we implicitly check that:
\begin{itemize}
    \item the index is within bounds $1 \leq k \leq \otimes\vec{s}$,
        as $F_{\vec{s}}$ is undefined outside the index space bounded by $\vec{s}$; and
    \item the index vector and the shape vector are of the same length,
        which means that selections evaluate scalars and not array sub-regions.
\end{itemize}

\paragraph{IMap}
%The most important rule for the semantics of \funclal{} is the rule for evaluating
%the \imap{}-construct.
In order to keep the \imap{} rule reasonably concise, we introduce two separate rules,
a rule \textsc{Gen}
for evaluating the generator bounds, and the main rule for \imap{}, the \textsc{Imap-Strict}-Rule:
\begin{mathpar}
    \inferrule[IMap-Strict]
        {
            S;\rho \vdash \semvar{e}{out} \evto S_1\sep \semvar{p}{out}
            \Rightarrow \tuple{\tuple{d_o}, \vec{\semvar{s}{out}}} \\
            S_1;\rho \vdash \semvar{e}{in} \evto S_2\sep \semvar{p}{in}
            \Rightarrow \tuple{\tuple{d_i}, \vec{\semvar{s}{in}}} \\
            %\otimes (\vec{\semvar{s}{out}} \plusplus \vec{\semvar{s}{in}}) < \omega \\
            \hat{S}_1 = S_2 \\
            \mathop{\forall}\limits_{i=1}^{n}
                \hat{S}_i;\rho \vdash g_i
                \evto \hat{S}_{i+1}\sep p_{g_i}
                \Rightarrow \bar{g}_i\\
            \text{FormsPartition}(\vec{\semvar{s}{out}}, \{\bar{g}_1, \dots, \bar{g}_n\})\\
            \bar{S}_{1} = \hat{S}_{n+1} \\
            \forall (i, \vec\imath) \in \text{Enumerate}(\vec{\semvar{s}{out}})
                \exists k:
                \left|{\begin{array}{l}
                    \vec{\imath} \in \bar{g}_k
                    \ \band\  \bar{g}_k = \Gen(x_k, \_, \_) \\
                    \bar{S}_{i}, p \mapsto \tuple{\tuple{d_o}, \vec{\imath}\,};
                        \rho, x_k \mapsto p \vdash e_k
                        \evto \bar{S}'_{i}\sep p_{\vec{\imath}} \\
                        \bar{S}'_i; \rho, x \mapsto p_{\vec\imath}
                        \vdash |x| \evto \bar{S}_{i+1}\sep p'_{\vec\imath}
                        \Rightarrow \tuple{\tuple{d_i}, \vec{\semvar{s}{in}}}
                       \end{array}}\right.\\
            \bar{S}_{\otimes\vec{\semvar{s}{out}}+1}, \rho \vdash
            \imap_1\ \semvar{p}{out}|\semvar{p}{in}\
                 \left\{\vec\imath
                        \mapsto p_{\vec\imath}\  |\
                        (\_,\vec\imath) \in \text{Enumerate}(\vec{\semvar{s}{out}})\right\}
            \evto S'\sep p
        }
        {
            S;\rho \vdash \imap\ \semvar{e}{out}|\semvar{e}{in}
            \ {\begin{cases}
                g_1: & e_1,\\
                \dots \\
                g_n: & e_n\\
            \end{cases}}
            \evto
            S'\sep p
        }
\end{mathpar}
\begin{mathpar}
    \inferrule[Gen]
        {
            S;\rho \vdash e_1 \evto S_1\sep p_1
            \Rightarrow \tuple{\tuple{n}, \vec{l}\,} \\
            S_1;\rho \vdash e_2 \evto S_2\sep p_2
            \Rightarrow \tuple{\tuple{n}, \vec{u}} \\
        }
        {
            S;\rho \vdash (e_1 \leq x < e_2)
            \evto S,p\mapsto \Gen(x, \vec{l}, \vec{u})\sep p
        }
\end{mathpar}
The \textsc{Gen}-rule introduces auxiliary values $\Gen(x, \vec{l}, \vec{u})$ which are
triplets that keep a variable name, lower bound and upper bound of a
generator together.
These auxiliary values are references only by the rule for \imap{}.

Evaluation of an \imap{} happens in three steps. First, we compute shapes
and generators, making sure that generators form a partition of
$\vec{\semvar{s}{out}}$ (FormsPartition is responsible for this).
Secondly, for every valid index defined by the frame shape (Enumerate
generates a set of offset-index-vector pairs), we find a
generator that includes the given index (denoted $\vec{\imath} \in \bar{g}_k$).
We evaluate the generator expression $e_k$, binding the generator variable $x_k$
to the corresponding index value and check that the result has the same shape as
$\semvar{p}{in}$.  Finally, we combine evaluated expressions for every index of
the frame shape into $\imap_1$ for further extraction of scalar values.

All missing rules, including built-in operations, conditionals and recursion through
the $letrec$-construct are straightforward adaptations of the standard rules.
They can be found in~\cite{lamom-sem}.
Formal definitions of helper functions, such as \text{AllSameShape},
will also be found there.

%%%%%%%%%%%%%%%%%%%%%%%%%%%%% END OF SEMANTICS %%%%%%%%%%%%%%%%%%%%%%%%%%%%%

\subsection{Infinite Arrays}

In order to support infinite arrays, we introduce the notion of infinity
in \funclal{}, and we allow infinities to appear in shape components.
Syntactically, this can be achieved by adding a symbol for infinity, as shown in
Fig.~\ref{fig:syntaxinf}.
For disambiguation, we refer to the extended version of \funclal{} as $\funclal^\infty$.
\begin{figure}[h!]
\begin{syntaxbox}[b]{.7\textwidth}{\funclal{} with cardinal infinity.\hfill
        extends \funclal{}}
\[
\begin{array}{ccll}
    c &::=& \cdots & \\
      &|& \infty & \text{(infinity constant)} \\
\end{array}
\]
\end{syntaxbox}
\caption{\label{fig:syntaxinf}The syntax of $\funclal^\infty$}
\end{figure}
Adding $\infty$ has several implications.
First of all, our built-in arithmetic needs to be extended.
We treat infinity in the usual way, applying the model commonly known
as a Riemann sphere.
That is:
\begin{mathpar}
    z + \infty = \infty \and
    z \times \infty = \infty \and
    \frac{z}{\infty} = 0 \and
    \frac{z}{0} = \infty
\end{mathpar}
The following operations are undefined:
\begin{mathpar}
    \infty + \infty \and
    \infty - \infty \and
    \infty \times 0 \and
    \frac{0}{0} \and
    \frac{\infty}{\infty}
\end{mathpar}

While these additions to the semantics are trivial, allowing infinity to
appear in shapes has a more profound impact on our semantics.
Our rule for \imap{}-constructs (\textsc{Imap-Strict}) forces the evaluation of all elements.
If our result shape contains infinity, this can no longer be done.
As we want to maintain a strict evaluation regime for
function applications in general,
we turn our \imap{}-construct into a lazy data-structure which does not immediately compute
its elements, but only does so when individual elements are being inspected.
For this purpose, we extend our
set of allowed values of our semantics with an \imap{}-closure:
\begin{mathpar}
    \closure{\imap\ \semvar{p}{out}|\semvar{p}{in}
             \ {\begin{cases}
                  \bar{g}_1: & e_1, \\
                  \dots\\
                  \bar{g}_n: & e_n
                \end{cases}}, \rho}
\end{mathpar}
The \imap{} closure contains pointers to frame
and element shapes ($\semvar{p}{out}$ and $\semvar{p}{in}$ correspondingly), the
list of partitions, where generators have been evaluated and the environment
in which the \imap{} shall be evaluated.
The overall idea is to update, in place,
this closure whenever individual elements are computed.
With this extension, we can now replace our strict \imap{}-rule by a lazy variant:

%%%%%%%%%%%%%%%%%%%%%%%%%%%%% BEGIN LAZY IMAP %%%%%%%%%%%%%%%%%%%%%%%%%%%%%
\begin{mathpar}
    \inferrule[IMap-Lazy]
        {
            S;\rho \vdash \semvar{e}{out} \evto S_1\sep \semvar{p}{out}
            \Rightarrow \tuple{\tuple{\_}, \vec{\semvar{s}{out}}}\\
            S_1;\rho \vdash \semvar{e}{in} \evto S_2\sep \semvar{p}{in}
            \Rightarrow \tuple{\tuple{\_}, \_}\\
            \hat{S}_1 = S_2 \\
            \mathop{\forall}\limits_{i=1}^{n}\hat{S}_i;\rho
              \vdash g_i \evto \hat{S}_{i+1}\sep p_{g_i} \Rightarrow \bar{g}_i \\
            %= \hat{S}_{n+1}(p_{g_i})\\
            \text{FormsPartition}(\vec{\semvar{s}{out}}, \{\bar{g}_1, \dots, \bar{g}_n\},)
        }
        {
            S;\rho \vdash \imap\ \semvar{e}{out}|\semvar{e}{in}
            \ {\begin{cases}
                    g_1: & e_1,\\
                    \dots \\
                    g_n: & e_n\\
               \end{cases}}
               \evto
               \hat{S}_{n+1},
               p\mapsto
               \closure{
                   \imap\ \semvar{p}{out}|\semvar{p}{in}
                   \ {\begin{cases}
                         \bar{g}_1: & e_1,\\
                         \dots \\
                         \bar{g}_n: & e_n\\
                      \end{cases}};
                   \rho}
               \sep p
        }
\end{mathpar}
We can see that the new rule for \imap{}-constructs, in essence,
performs a subset of what the strict rule from the previous
section does. It still forces the result shapes,
it still computes the boundaries of the generators,
and it checks the validity of the overall generator set.
Once these computations have been done, further element computation is delayed and
an imap{}-closure is created instead.

The actual computation of elements is triggered upon element selection.
Consequently, we need a second selection rule which can deal with \imap{} closures
in the array argument position:
\begin{mathpar}
    \inferrule[Sel-lazy-imap]
        {
            S;\rho \vdash i \evto S_1\sep p_i
            \Rightarrow
            \tuple{\tuple{\_}, \vec{v}\,} \\
            S_1;\rho \vdash a \evto S_2\sep p_a
            \Rightarrow \closure{\imap\ \semvar{p}{out}|\semvar{p}{in}\
                                {\begin{cases}
                                    \bar{g}_1 & e_1\\
                                    \dots\\
                                    \bar{g}_n & e_n\\
                                 \end{cases}}, \rho'}\\
            S_2 (\semvar{p}{out}) = \tuple{\tuple{m},\_}\\
            %S_2 (\semvar{p}{in}) = \tuple{\tuple{r}, \_}\\
            %l = m + r \\
            %\vec{\imath} = \tuple{i_1, \dots, i_m} \\
            %\vec{\jmath} = \tuple{i_{m+1}, \dots, i_l} \\
            (\vec\imath, \vec\jmath) = \text{Split}(m, \vec{v}\,) \\
            \exists k: \vec{\imath} \in \bar{g}_k\\
            \bar{g}_k = \Gen(x_k, \_, \_) \\
            S_2, p \mapsto \Exp{\vec{\imath}};\rho', x_k \mapsto p
            \vdash e_k \evto S_3\sep p_{\vec{\imath}} \\
            S_3; \rho', x \mapsto p_{\vec{\imath}}
            \vdash x.\Exp{\vec\jmath} \evto S_4\sep p\\
            S_5 = \text{UpdateIMap}(S_4, p_a, \vec\imath, p_{\vec\imath})
            % FIXME what if p_i = <>?  Introduce the same Split function as in
            % imap_1
        }
        {
            S;\rho \vdash a.i \evto S_5\sep p
        }
\end{mathpar}
Selections into \imap{}-closures happen at indices that are of the same
length as the concatenation of the \imap{} frame and cell shapes.  This means that
the index the \imap{}-closure is being selected from has to be split
into frame and cell sub-indices: $\vec\imath$ and $\vec\jmath$ correspondingly.  Given that
$\bar{g}_k$ contains $\vec\imath$, we evaluate $e_k$ with $x_k$ being bound
to $\vec\imath$.  As this value may
be non-scalar, we evaluate a selection into it at $\vec\jmath$.
Finally, the evaluated generator expression is saved within the \imap{}
closure.  This step is performed by the helper function $\text{UpdateIMap}$,
which splits the $k$-th partition into a single-element partition containing
$\vec\imath$ with the computed value $p_{\vec\imath}$, and further partitions
covering the remaining indices of $\bar{g}_k$ with the expression $e_k$.
For more details see~\cite[Sec.~2.1.1]{lamom-sem}.

%%%%%%%%%%%%%%%%%%%%%%%%%%%%% END LAZY IMAP %%%%%%%%%%%%%%%%%%%%%%%%%%%%%

With this, we can define and use infinite arrays in an overall strict setting.
 Let us consider the definitions
of the infinite array of natural numbers in $\funclal^\infty$ on the left and Haskell-like
definition on the right:

\noindent
\begin{tabular}{@{}p{.49\textwidth}p{.49\textwidth}}
{\begin{lstlisting}[mathescape, style=nonumbers]
nats $\equiv$ imap [$\infty$] { _(iv): iv.[0]
\end{lstlisting}
}
&
{\begin{lstlisting}[mathescape, style=nonumbers, language=Haskell]
nats = 0: map (+1) nats
\end{lstlisting}
}
\end{tabular}

Both versions define an object that delivers the value $n$ when being selected at any
index $n$.  Both definitions provide a data structure whose computation unfolds
in a lazy fashion.
The main difference is that the Haskell definition enforces a left-to-right unfolding
of the list.
Whenever an element $n$ is selected, the entire spine of
the list, up to the $n$-th element, has to be in place.
In the $\funclal^\infty$ case, any element can be computed directly.
The actual access time as well as the storage demand depend on how the
\textsc{Imap-Lazy}-rule is being implemented.
In particular, it depends on how the \imap{}-closure is being updated
by an implementation of the UpdateIMap operation.

The above comparison demonstrates the fundamental difference between a data-parallel
programming style and a list-based, inherently recursive programming style.
Even if the former is mimicked by the latter using list comprehensions,
\eg{} $\progvar{nats} = [i\ |\ i \mathrel{{<}{-}} [0..]]$, the idiom $[0..]$ boils down
to a recursive construction of the spine of the list.

Having observed this fundamental difference, one may wonder
if these kinds of Haskell-like recursive definitions
are possible in $\funclal^\infty$ at all?

\subsection{Recursive Definitions}

It turns out that the lazy \imap{}, together with the \emph{letrec}
construct, allows for recursive definitions of arrays.
A recursive definition of the natural numbers, including 0, can be defined
in $\funclal^\infty$ by:
\begin{lstlisting}[mathescape, style=nonumbers]
letrec nats = imap [$\infty$] { [0] <= iv < [1]: 0,
                        $\;$ [1] <= iv < [$\infty$]: nats.(iv$\dot{-}$[1]) + 1 in nats
\end{lstlisting}

% \fixme{If we have infinities in shape vectors, can we obtain
% any performance (or analysis) benefits from the frame/cell split?
% E.g., if the infinities appear only in the frame, but the cell
% is finite. Or vice versa. If so, can we give a suitable example
% of same?}
%
% I don't see how the frame/cell shape would help here.

The interesting question here is whether the semantics defined thus
far ensures that all
elements of the array nats are actually being inserted into one and the same \imap{}-closure.
For this to happen, we need the environment of the \imap{}-closure to
map nats to itself, and we need the selection within the body
of the imap{} to modify the closure from which it is selecting.
While the latter is given through the \textsc{Sel-Lazy-Imap}-rule,
the former is achieved through the rule for letrec-constructs. For \funclal{}, we have:
\begin{mathpar}
    \inferrule[Letrec]
        {
            S_1 = S, p \mapsto \bot \\
            \rho_1 = \rho, x \mapsto p \\
            S_1;\rho_1 \vdash e_1 \evto S_2\sep p_2 \\
            S_3 = S_2[p_2/p] \\
            S_3;\rho,x\mapsto p_2 \vdash e_2 \evto S_4\sep p_r
        }
        {
            S;\rho \vdash \letrec{x = e_1}{e_2} \evto S_4\sep p_r
        }
\end{mathpar}
%\todo[inline]{Point to Joe's paper on letrecs.}
where $S[p_2/p]$ denotes substitution of the $x\mapsto p$ bindings
inside of the enclosed environments with $x\mapsto p_2$, where $x$ is any legal
variable name.
This substitution is key for creating the circular reference in the \imap{}-closure
from the example above.

In conclusion, the above recursive specification
denotes an array with the same elements as the data-parallel specification
from the previous section.
In contrast to data-parallel version, this specification behaves much more
like the recursive, Haskell-like
version; the computation of individual elements can no longer happen directly.
Since there is an encoded dependency between an element and its predecessor, the first access
to an element at index $n$, in this variant, will trigger the computation of all elements
from $0$ up to $n$.
The implementation of the UpdateIMap operation on \imap{}-closures determines
how these numbers are stored in memory and, consequently, how
efficiently they can be accessed.

The availability of direct indexes makes it possible to encode an
arbitrary order for the recursion. Consider the following example:
\begin{lstlisting}[mathescape, style=nonumbers]
letrec a = imap [10] { [9] <= iv < [10]: 9,
                       [0] <= iv < [9]:  a.(iv$\dot{+}$[1])-1 in a
\end{lstlisting}
Selection of the 9th element
can be evaluated in one step.  In case of lists, the selection request always
starts at the beginning of the list. Hence, to obtain the same
performance, some optimisation of the list case is required.

\subsection{List Primitives in the Array Setting}

We have enabled two features that are inherent with lists,
but that are usually not supported
in an array setting: recursively defined data-structures and infinite arrays.
All that is required to achieve this is a recursion-aware, lazy semantics of the
\imap{}-construct and the inclusion of an explicit notion of infinity.
With these extensions, the key primitives for lists, $head$, $tail$, and $cons$ can be defined as
\begin{lstlisting}[mathescape, style=nonumbers]
head $\equiv$ $\lambda$a.a.[0]
tail $\equiv$ $\lambda$a.imap |a|$\dot{-}$[1] { _(iv): a.([1]$\dot{+}$iv)
cons $\equiv$ $\lambda$a.$\lambda$b.imap [1]$\dot{+}$|b| { [0] <= iv < [1]: a,
                            $\,$[1] <= iv < [1]$\dot{+}$|b|: b.(iv$\dot{-}$[1])
\end{lstlisting}

More complex list-like functions can be defined on top of these.
An example is concatenation:
\begin{lstlisting}[mathescape, style=nonumbers]
letrec (++) = $\lambda$a.$\lambda$b.if |a|.[0] = 0 then b
                    else cons (head a) ((tail a) ++ b) in (++)
\end{lstlisting}
In case $a$ is infinite, however, the above definition of concatenation
is unsatisfying. The strict nature of \funclal{} will force $\progvar{tail}\ a$ forever as
$\shp{a}.[0] = 0$ never yields \true{}.
The way to avoid this is to shift the case distinction into the lazy \imap{}
construct:
\begin{lstlisting}[mathescape, style=nonumbers]
(++) $\equiv$ $\lambda$a.$\lambda$b.imap |a|$\dot{+}$|b| { [0] <= iv < |a|:     a.iv,
                            $\,$|a| <= iv < |a|$\dot{+}$|b|: b.(iv$\dot{-}$|a|)
\end{lstlisting}

As we have seen earlier, \funclal{} enables the typical constructions of
recursive definitions of infinite vectors well-known from the realm of lists
such as list of ones, natural numbers or fibonacci sequence.

Having a unified interface for arrays and lists enables programmers to switch
the algorithmic definitions of individual arrays from recursive to data-parallel
styles without modifying any of the code that operates on them.

However, such a unification comes at a price: we have to support a lazy version of
the \imap{}-construct.  As a consequence, we conceptually lose the
advantage of $O(1)$ access.
Despite \funclal{} offering many opportunities for compiler
optimisations like pre-allocating arrays and potentially enforcing strictness on
finite, non-recursive \imap{}s, one may wonder at this point how much \funclal{}
differs from a lazy array interface in a lazy, list-based language such as
Haskell?

\section{Transfinite Arrays}

We now investigate to what extent $\funclal^\infty$ adheres to
the key properties of array programming ---
array algebras and array equalities.

\subsection{\label{sec:transfinite:alg}Algebraic Properties}

Array-based operations offer a number of beneficial algebraic properties.
Typically, these properties manifest themselves as
universally valid equalities which, once established, improve
our thinking about algorithms and their implementations, and give
rise to high-level program transformations.
We define equality between two non-scalar arrays $a$ and $b$ as
\[
    a == b
    \Longleftrightarrow
    |a| = |b|
    \band \mathop{\forall} \progvar{iv} < |a|:
    a.\progvar{iv} = b.\progvar{iv}
\]
that is, we demand equality of the shapes and equality
of all elements.
The demand for equality of shapes recursively implies equality in dimensionality
and the extensional character of this definition through the use of array selections
ensures that we can reason about equality on infinite arrays as well.

Arrays give rise to many algebras such as Theory of
Arrays~\cite{more-arrayth}, Mathematics of Arrays~\cite{LMRMullin:moa},
and Array Algebras~\cite{Jenkins88}.  Most of the
developed algebras differ only slightly, and the set of equalities that
are ultimately valid depends on some fundamental choices, such as the ones we
made in the beginning of the previous section.  At the core of these
equalities is the ability to change the shape of arrays in a
systematic way without losing any of their data.

An equality from~\cite{jenkins99} that plays a key role in consistent shape
manipulations is:
\begin{equation}\label{eqn:reshape}
    \progvar{reshape}\ |a|\ (\progvar{flatten}\ a)\ ==\ a
\end{equation}
where \progvar{flatten} maps an array recursively into a vector by
\emph{concatenating}
its sub-arrays in a row-major fashion and \progvar{reshape} performs the dual operation
of bringing a row-major linearisation back into multi-dimensional form.
These operations can be defined in $\funclal^\infty$ as
\begin{lstlisting}[mathescape, style=nonumbers]
flatten $\equiv$ $\lambda$a.imap [count a] { _(iv): a.(o2i iv.[0] |a|)
reshape $\equiv$ $\lambda$shp.$\lambda$a.imap shp { _(iv): (flatten a).[i2o iv shp]
\end{lstlisting}
where \progvar{count} returns the product of all shape components
and \progvar{o2i} and \progvar{i2o} translate offsets into indices and vice
versa, respectively. These operations effectively implement
conversions from mixed-radix systems into natural numbers using multiplications and additions
and back using division and remainder operations.

The above equality states that any array $a$ can be brought into flattened form and,
subsequently be brought back to its original shape.
For arrays of finite shape $s$, this follows directly from the fact
that $\progvar{o2i}\ ( \progvar{i2o}\ \progvar{iv}\ s)\ s = \progvar{iv}$
for all legitimate index vectors \progvar{iv} into the shape $s$.

If we want Eq.~\ref{eqn:reshape} to hold for all arrays in $\funclal^\infty$,
we need to show that the above equality also holds for arrays with infinite axes.
Consider an array of shape $s = [2, \infty]$.
For any legal index vector $[1, n]$ into the shape $s$, we obtain:
\begin{align*}
o2i\ (i2o\ [1,n]\ [2,\infty])\ [2,\infty])
 &= o2i\ (\infty \cdot 1 + n) \ [2,\infty] \\
 &= o2i\ \infty\ [2,\infty] \\
 &= [\infty \ /\ \infty,\ \infty \ \%\ \infty]
\end{align*}
which is not defined.
We can also observe that all indices $[1, n]$ are effectively mapped into the same offset: $\infty$ which is
not a legitimate index into any array in $\funclal^\infty$.
This reflects the intuition that the concatenation of two infinite vectors
effectively looses access to the second vector.

The inability to concatenate infinite arrays also makes the following equality
fail:
\begin{equation}\label{eqn:takedrop}
    \progvar{drop}\ |a|\ (a\plusplus b) == b
\end{equation}
where $a$ and $b$ are vectors and $\progvar{drop}\ s\ x$ removes first $s$
elements from the left.  The reason is exactly the same: given that $|a| =
[\infty]$ and $b$ is of finite shape $[n]$, the shape of the concatenation
is $[\infty + n] = [\infty]$, and drop of $|a|$ results in an empty vector.

%In contrast, the validity of the above equation is less clear in case the shape of an array contains infinities.
%For the case where we have $a\ \equiv\ imap\ [\infty,2]\ \{\ \_(iv):\ iv.[1]\ =\ [[0,1],[0,1],[0,1],\ldots]$ one
%may hope to obtain
%\begin{equation*}
%  \begin{split}
%    reshape\ |a|\ (flatten\ a)&
%    =\ reshape\ [\infty,2] (flatten\ [[0,1],[0,1],[0,1],\ldots])\\
%    & =\ reshape\ [\infty,2]\ imap\ [\infty]\ \{\ \_(iv):\ iv.[0]\ \%\ 2\\
%    & =\ reshape\ [\infty,2]\ [0,1,0,1,0,1,\ldots]\\
%    & =\ imap\ [\infty,2]\ \{\ \_(iv):\ iv.[1]\\
%    & =\ a \qquad .
%  \end{split}
%\end{equation*}
%
%If we look at $a$ being defined by $a\ \equiv\ imap\ [2,\infty]\ \{\ \_(iv):\ iv.[0]\ =\ [[0,1,2,3,\ldots], [0,1,2,3,\ldots]]$
%we would need to be able to describe the flattened version of $a$ as an \imap{}-construct.
%Unfortunately, that requires accessing elements \emph{beyond} infinitely many elements which the approach
%proposed so far does not allow us to do.
%Not being able to express the flattened version of this array means we will not be able to prove equation~(\ref{eqn:reshape})
%for infinite arrays as we have introduces them so far in general.
%We deem this a price to high to pay.

Clearly, $\funclal^\infty$ as presented so far is not strong enough to maintain
universal equalities such as Eq.~\ref{eqn:reshape} or~\ref{eqn:takedrop}.
Instead, we have to find a way that enables us to represent sequences of 
infinite sequences that can be distinguished from each other.
% One of the key contributions of this paper is the suggestion to reestablish these
% equalities by using ordinals for indices and shape descriptions.

\subsection{Ordinals}

When numbers are treated in terms of cardinality,
they describe the number of elements in a set.
Addition of two cardinal numbers $a$ and $b$ is defined as
a size of a union of sets of $a$ and $b$ elements.  This
notion also makes it possible to operate with infinite numbers, where the
number of elements in an infinite set is defined via bijections.
As a result, differently constructed infinite sets may end up having the same
number of elements.  For example, if there exists a bijection from
$\mathbb{N}\times\mathbb{N}$ into $\mathbb{N}$, the cardinality
of both sets is the same.

When studying arrays, treating their shapes and indices using cardinal numbers
is an oversimplification, because arrays have richer
structure.  Arrays are collections of ordered elements, where the order is
established by the indices.  Ordinal numbers, as introduced by G. Cantor
in 1883, serve exactly this purpose --- to ``label'' positions of objects within
an ordered collection.  When collections are finite, cardinals and ordinals can
be used interchangeably, as we can always count the labels.  Infinite
collections are quite different in that regard: despite being of the same size,
there can be many non-isomorphic well-orderings of an infinite collection.
For example, consider two infinite arrays of shapes $[\infty, 2]$ and $[2, \infty]$.
Both of these have infinitely many elements, but they differ in their structure.
From a row major perspective, the former is an infinite sequence of pairs,
whereas the latter are two infinite sequences of scalars.
Ordinals give a formal way of describing such different well-orderings.

First let us try to develop an intuition for the concept of ordinal numbers
and then we give a formal definition.
Consider an ordered sequence of natural numbers:
\(
    0 < 1 < 2 < \cdots.
\)
These are the first ordinals.  Then, we introduce a number called $\omega$ that
represents the limit of the above sequence:
\(
    0 < 1 < 2 < \cdots < \omega.
\)
Further, we can construct numbers beyond $\omega$ by putting a ``copy'' of
natural numbers ``beyond'' $\omega$:
\[
    0 < 1 < 2 < \cdots \omega < \omega+1 < \omega+2 < \cdots < \omega+\omega
\]
For the time being, we treat operations such as $\omega+n$ symbolically.
The number $\omega+\omega$
which can be also denoted as $\omega\cdot2$ is the second limit ordinal
that limits any number of the form
$\omega+n, n \in \mathbb{N}$.  Such a procedure of
constructing limit ordinals out of already constructed smaller ordinals
can be applied recursively.  Consider a sequence of $\omega\cdot n$
numbers and its limit:
\[
    0 < \omega < \omega\cdot2 < \omega\cdot3 < \cdots < (\omega\cdot\omega = \omega^2)
\]
and we can carry on further to $\omega^n$, $\omega^\omega$, \etc{}
Note though that in the interval from $\omega^2$ to $\omega^3$ we have
infinitely many limit ordinals of the form:
\[
    \omega^2< \omega^2+\omega < \omega^2+\omega\cdot2 < \cdots < \omega^3
\]
and between any two of these we have a ``copy'' of the natural numbers:
\[
    \omega^2+\omega < \omega^2+\omega+1 < \cdots < \omega^2+\omega\cdot 2
\]

\subsubsection{Formal definitions}

A totally ordered set $\tuple{A, <}$ is said to be well ordered
(or have a well-founded order) if and only if every nonempty subset of
$A$ has a least element~\cite{settheory}.
Given a well-ordered set $\tuple{X, <}$ and $a \in X$,
\(
    X_a \stackrel{\text{def}}{=} \{x \in X | x < a\}
\).
An ordinal is a well-ordered set $\tuple{X, <}$, such that:
\(
    \forall a \in X: a = X_a
\).
As a consequence, if $\tuple{X, <}$ is an ordinal then $<$ is equivalent
to $\in$.
%Isomorphism?
Given a well-ordered set $A = \tuple{X, <}$ we define an ordinal that this set is
isomorphic to as $Ord(A, <)$.
Given an ordinal $\alpha$, its successor is defined to be $\alpha \cup
\{\alpha\}$.  The minimal ordinal is $\emptyset$ which is denoted with $0$.  The
next few ordinals are:
\[
\begin{array}{lclcl}
    1 &=& \{0\}   &=& \{\emptyset\} \\
    2 &=& \{0,1\} &=& \{\emptyset, \{\emptyset\}\} \\
    3 &=& \{0,1,2\} &=& \{\emptyset, \{\emptyset\}, \{\emptyset, \{\emptyset\}\}\} \\
    &&\cdots&&
\end{array}
\]

A limit ordinal is an ordinal that is greater than zero that is not a successor.
The set of natural numbers $\{0, 1, 2, 3, \dots\}$ is the smallest limit ordinal
that is denoted $\omega$.
We use $\lo\ x$ to denote that $x$ is a limit ordinal.

\subsubsection{Arithmetic on Ordinals}

\paragraph{Addition} Ordinal addition is defined as
$\alpha + \beta = Ord(A, <_A)$, where
$A = \{0\}\times\alpha \mathrel{\cup} \{1\}\times\beta$ and $<_A$ is the
lexicographic ordering on $A$.
Ordinal addition is associative but \emph{not} commutative.  As an example
consider expressions $2 + \omega$ and $\omega + 2$.  The former can be seen
as follows:
\(
    0 < 1 < 0' < 1' < \cdots
\),
which after relabeling is isomorphic to $\omega$.  However, the latter can
be seen as:
\(
    0 < 1 < \cdots < 0' < 1'
\),
which has the largest element $1'$, whereas $\omega$ does not.  Therefore
$2 + \omega = \omega < \omega + 2$.  We have used $0'$, $1'$ to indicate the
right hand side argument of the addition.

\paragraph{Subtraction} Ordinal subtraction can be defined in two ways,
as partial inverse of the addition on the left and on the right.
For left subtraction, which will be used by default throughout this paper unless
otherwise specified,  $\alpha - \beta$ is defined when $\beta \leq \alpha$,
as: $\exists\xi: \beta + \xi = \alpha$.  As ordinal addition is
left-cancelative ($\alpha + \beta = \alpha + \gamma \implies \beta = \gamma$),
left subtraction always exists and it is unique.

Right subtraction is a bit harder to define as:
\begin{itemize}
    \item it is not unique: $1 + \omega = 2 + \omega$ but $1 \not=2$;
        therefore $\omega -_{R} \omega$ can be any number that is less
        than $\omega$: $\{0, 1, 2, \dots\}$.
    \item even if $\beta < \alpha$, the difference $\alpha - \beta$ might
        not exist.  For example: $42 < \omega$; however, $\omega -_{R} 42$
        does not exist as $\not\exists\xi: \xi + 42 = \omega$.
\end{itemize}
Despite those difficulties, right subtraction can be useful at times and can
be defined for $\alpha -_{R} \beta$:
\[
    \min \{\xi: \xi + \beta = \alpha\}
\]

\paragraph{Multiplication} Ordinal multiplication
$\alpha \cdot \beta = Ord(A, <_A)$ where $A = \alpha \times \beta$
and $<_A$ is the lexicographic ordering on $A$.  Multiplication is associative
and left-distributive to addition:
\[
    \alpha \cdot (\beta + \gamma) = (\alpha\cdot \beta) + (\alpha \cdot \gamma)
\]
However, multiplication is not commutative and is not distributive on the
right: $2\cdot\omega = \omega < \omega\cdot2$ and $(\omega+1)\cdot \omega =
\omega\cdot\omega < \omega\cdot\omega+\omega$.

\paragraph{Exponentiation} Exponentiation can be defined using transfinite
recursion: $\alpha^0 = 1, \alpha^{\beta+1} = \alpha^\beta\cdot\alpha$ and for limit
ordinals $\lambda$: $\alpha^\lambda = \bigcup\limits_{0 < \xi < \lambda}
\alpha^\xi$.

\paragraph{$\epsilon$-ordinals}
Using ordinal operations we can construct the following hierarchy of ordinals:
$0,1,\dots,\omega,\omega+1,\dots,\omega\cdot2,\omega\cdot2+1,\dots, \omega^2,
\dots,\omega^3,\dots\omega^\omega,\dots$.  The smallest ordinal for which
$\alpha = \omega^\alpha$ is called $\epsilon_0$.  It can also be seen as a
limit of the following $\omega^\omega, \omega^{\omega^\omega}, \dots,
\omega^{\omega^{\dots}}$.

\subsubsection{Cantor Normal Form}
For every ordinal $\alpha < \epsilon_0$ there are unique $n,p < \omega,
\alpha_1 > \alpha_2 > \cdots > \alpha_n$ and $x_1, \dots, x_n \in \omega
\setminus \{0\}$ such that $\alpha > \alpha_1$ and $\alpha =
\omega^{\alpha_1}\cdot x_1 + \cdots + \omega^{\alpha_n}\cdot x_n + p$.
Cantor Normal Form makes provides a standardized way of writing
ordinals. It uniquely represents each ordinal as a finite sum of
ordinal powers, and can be seen as an $\omega$ based polynomial.
This can be used as a basis for an efficient implementation
of ordinals and their operations.

\subsection[Lambda-omega: Adding Ordinals to lambda-alpha]%
{\label{sect:trans:lom}\funclom{}: Adding Ordinals to \funclal{}}

The key contribution of this paper is the introduction of \funclom{}, a variant
of \funclal{}, which use ordinals as shapes and indices of arrays and which
reestablishes global equalities in the context of infinite arrays.

Before revisiting the equalities, we look at the changes to \funclal{} that are
required to support transfinite arrays.
Syntactically, to introduce ordinals in the language, we make two minor
additions to \funclal{}. Firstly, we add ordinals\footnote{
    Technically, we support ordinal values only up to $\omega^\omega$, as
    ordinals are constructed using the constant $\omega$ and $+$, $-$, $*$, $/$
    and $\%$ operations (no built-in ordinal exponentiation).}
as scalar constants.  Secondly, we add a built-in operation, \lo{},
which takes one argument and returns \true{} if the argument is a
limit ordinal and \false{} otherwise.  For example: $\lo\ \omega$ reduces
to \true{} and $\lo\ (\omega + 21)$ reduces to \false{}.

\begin{figure}[h!]
\begin{syntaxbox}[b]{.7\textwidth}%
                    {\funclal{} with ordinals\hfill extends \funclal{}}
\[
\begin{array}{ccll}
    e &::=& \cdots & \\
      &|& \lo & \text{(limit ordinal predicate)} \\

    c &::=& \cdots & \\
      &|& \omega,\omega+1,\dots & \text{(ordinals)} \\
\end{array}
\]
\end{syntaxbox}
\caption{\label{fig:syntaxomega}The syntax of \funclom{}.}
\end{figure}

Semantically, it turns out that all core rules can be kept unmodified
apart from the aspect that all helper functions, arithmetic, and relational operations
now need to be able to deal with ordinals instead of natural numbers.
In particular, the semantic for lazy \imap{}s as developed for $\funclal^\infty$
can be used unaltered, provided that all helper functions involved such as for splitting
generators \etc{} are expanded to cope with ordinals.

\subsection{Array Equalities Revisited}

With the support of Ordinals in \funclom{}, we can now revisit our equalities
Eq.~\ref{eqn:reshape} and~\ref{eqn:takedrop}.  Let us first look at the counter
example for Eq.~\ref{eqn:reshape}: from Section~\ref{sec:transfinite:alg}: With
an array shape $s = [2, \omega]$ and a legal index vector into $s$
$[1, n]$, we now obtain:
{
\begin{align*}
o2i\ (i2o\ [1,n]\ [2,\omega])\ [2,\omega])
 &= o2i\ (\omega + n) \ [2,\omega] \\
 &= [(\omega + n)\ /\ \omega,\ (\omega + n)\ \%\ \omega] \\
 &= [1,\ n]
\end{align*}
}
The crucial difference to the situation from $\funclal^\infty$ in Section~\ref{sec:transfinite:alg}
here is the ability to divide $(\omega + n)$ by $\omega$ and
to obtain a remainder, denoted by $\%$, of that division as well.
By means of induction over the length of the shape and index vectors this
equality can be proven to hold for arbitrary shapes in \funclom{}, and, based on
this proof, Eq.~\ref{eqn:reshape} can be shown as well.

In the same way as the arithmetic on ordinals is key to the proof of Eq.~\ref{eqn:reshape},
it also enables the proof of Eq.~\ref{eqn:takedrop} for arbitrary
ordinal-shaped vectors\footnote{
    Eq.~\ref{eqn:takedrop} can be generalised and shown to hold in the multi-dimensional case, provided that
    $\plusplus$ and \progvar{drop} operate over the same axis.}
$a$ and $b$, with the definition of $\plusplus$ from the
previous section and \progvar{drop} being defined as:
\begin{lstlisting}
drop $\equiv$ $\lambda$s.$\lambda$a. imap |a|$\dot{-}$s { [0] <= iv < |a|$\dot{-}$s: a.(s$\dot{+}$iv)
\end{lstlisting}
After inlining $\plusplus$ and \progvar{drop},
the left hand side of Eq.~\ref{eqn:takedrop} can be rewritten as:

\begin{lstlisting}
letrec ab = imap |a|$\dot{+}$|b| { [0] <= jv < |a|: a.jv,
                        $\,$   |a| <= jv < |a|$\dot{+}$|b|: b.(jv$\dot{-}$|a|) in

imap |ab|$\dot{-}$|a| { [0] <= iv < |ab|$\dot{-}$|a|: ab.(|a|$\dot{+}$iv)
\end{lstlisting}

Consider the shape of the goal expression of the letrec.  According to the
semantics of the shape of an \imap{}, we get: $|\progvar{ab}|\dot{-}|a|$.  The
shape of \progvar{ab} is $|a|\dot{+}|b|$.  According to ordinal arithmetic:
$(|a|\dot{+}|b|)\dot{-}|a|$ is $|b|$.  Therefore the shapes of right-hand and
left-hand sides of the goal expressions are the same.

Let us rewrite the last \imap{} as:
\begin{lstlisting}
imap |b| { [0] <= iv < |b|: ab.(|a|$\dot{+}$iv)
\end{lstlisting}

Consider now selections into \progvar{ab}.  All the selections into \progvar{ab}
will happen at indices that are greater than $a$.  This is because all the legal
\progvar{iv} in the \imap{} are from the range $[0]$ to $|b|$.

According to the semantics of selections into \imap{}s,
$\progvar{ab}.(|a|\dot{+}\progvar{iv})$ will select from the second partition of
the \imap{} that defines \progvar{ab}, and will evaluate to:
$b.((|a|\dot{+}\progvar{iv})\dot{-}|a|)$.  According to ordinal arithmetic,
$(|a|\dot{+}\progvar{iv})\dot{-}|a|$ is identical to \progvar{iv}, therefore we
can rewrite the previous \imap{} as:
\begin{lstlisting}
imap |b| { [0] <= iv < |b|: b.iv
\end{lstlisting}
As it can be seen, this is an identity \imap{}, which is extensionally
equivalent to $b$.

\section{Examples}

\paragraph{Transfinite tail}
As explained in Section~\ref{sect:trans:lom}, the shift from natural numbers
to ordinals as indices in \funclom{} implies corresponding extensions
of the built-in arithmetic operations.
As these operations lose key properties, such as commutativity, once arguments
exceed the range of natural numbers, we need to ensure that function definitions
for finite arrays extend correctly to the transfinite case.

As an example, consider the definition of \progvar{tail} from the previous section:
\begin{lstlisting}[mathescape, style=nonumbers]
tail $\equiv$ $\lambda$a.imap |a|$\dot{-}$[1] {_(iv): a.([1]$\dot{+}$iv)
\end{lstlisting}
For the case of finite vectors, we can see that a vector shortened by one element
is returned, where the first element is missing and all elements have been shifted to the left by one element.

Let us assume we apply tail to an array $a$ with $|a| = [\omega]$.
The arithmetic on ordinals gives us a return shape of $[\omega]\dot{-}[1] = [\omega]$.
That is, the tail of an infinite array is the same size
as the array itself, which matches our common intuition when applying tail to infinite lists.
The elements of that infinite list are those of $a$, shifted by one element to the right,
which, again, matches our expected interpretation for lists.

Now, assume we have $|a| = [\omega+42]$, which means that
$(\progvar{tail}\ a) . [\omega]$ should be a valid expression.
For the result shape of $\text{tail}\ a$, we obtain $[\omega+42]\dot{-}[1] = [\omega+42]$.
A selection $(\text{tail}\ a) . [\omega]$ evaluates to $a . ([1]\dot{+}[\omega]) = a . [\omega]$.
This means that the above definition of the tail shifts all the elements at indices smaller
than $[\omega]$ one left, and leaves all the other unmodified.
While this may seem counter-intuitive at first, it actually only means that
\progvar{tail} can
be applied infinitely often but will never be able to reach ``beyond'' the first limit.

Finally, observe that the body of the \imap{}-construct in the definition of
\progvar{tail}
uses $[1]\dot{+}\progvar{iv}$ is an index expression, not $\progvar{iv}\dot{+}[1]$.
In the latter case, the tail function would behave differently
beyond $[\omega]$: it would attempt to shift elements to the left.  However,
this would make the overall definition faulty.  Consider again the case
when $|a| = [\omega+42]$: the shape of the result would be $|a|$, which would mean
that it would be possible to index at position $[\omega+41]$, triggering
evaluation of $a.([\omega+41]\dot{+}[1])$ and consequently, producing
an \emph{index error}, or out-of-bounds access into $a$.

\paragraph{Zip}
Let us now define zip of two vectors that produces a vector of tuples.
Consider a Haskell definition of zip function first:
\begin{lstlisting}[mathescape, style=nonumbers, language=Haskell]
zip (a:as) (b:bs) = (a,b) : zip as bs
zip _      _      = []
\end{lstlisting}
The result is computed lazily, and the length of
the resulting list is a minimum of the lengths of the arguments.  Like
concatenation, a literal translation into \funclom{} is possible, but
it has the same drawbacks, \ie{} it is restricted to arrays whose shape
has no components bigger than $\omega$.

%\begin{lstlisting}[mathescape, style=nonumbers]
%zip$_1$ a b $\equiv$ $\lambda$a.$\lambda$b.
%           if |a| == [0] or |b| == [0] then []
%           else (imap [2] {[0] <= iv < [1]: head a, [1] <= iv < [2]: head b)
%                ++
%                (zip (tail a) (tail b))
%\end{lstlisting}

A better version of zip that can be applied to arbitrary transfinite arrays looks as follows:
%pair a b $\equiv$ $\lambda$a.$\lambda$b.
%            imap [2] {[0] <= jv < [1]: a iv,
%                      [1] <= jv < [2]: b iv
%
\begin{lstlisting}[mathescape, style=nonumbers]
zip $\equiv$ $\lambda$a.$\lambda$b.imap (min |a| |b|)|[2] {_(iv): [a.iv, b.iv]
\end{lstlisting}
Here, we use a constant array in the body of the \imap{}. This
forces evaluation of both arguments, even if only one of them is selected.
This can be changed by replacing the constant array with an \imap{}:
\begin{lstlisting}[mathescape, style=nonumbers]
zip $\equiv$ $\lambda$a.$\lambda$b.imap (min |a| |b|)|[2] {_(iv): imap [2] { [0] <= jv < [1] a.iv,
                                                      [1] <= jv < [2] b.iv
\end{lstlisting}
which can be fused in a single \imap{} as follows:
\begin{lstlisting}[mathescape, style=nonumbers]
zip $\equiv$ $\lambda$a.$\lambda$b.letrec s = (min |a| |b|).[0] in
            imap [s,2] { [0,0] <= iv < [s, 1]: a.[iv.[0]],
                         [0,1] <= iv < [s, 2]: b.[iv.[0]]
\end{lstlisting}

\paragraph{Data Layout and Transpose}
A typical transformations in stream programming is
changing the granularity of a stream and joining multiple streams.  In
\funclom{}, these transformations can be expressed by manipulating the shape
of an infinite array.  Consider changing the granularity of a stream $a$ of shape
$[\omega]$ into a stream of pairs:
\begin{lstlisting}[mathescape, style=nonumbers]
imap (|a|$\dot{/}$[2])|[2] { _(iv): [a.[2*iv.[0]], a.[2*iv.[0]+1]]
\end{lstlisting}
or we can express the same code in a more generic fashion:
\begin{lstlisting}[mathescape, style=nonumbers]
($\lambda$n.reshape ((|a|$\dot{/}$[n])++[n]) a) 2
\end{lstlisting}
This code can operate on the streams of transfinite length, as well.
If we envision compiling such a program into machine code, the infinite
dimension of an array can be seen as a time-loop, and the
operations at the inner dimension seen as a stream-transforming function.
Such granularity changes are often essential for making good use of (parallel)
hardware resources, \eg{} FPGAs.

Transposing a stream makes it possible to introduce synchronisation.  Consider
transforming a stream $a$ of shape $[2, \omega]$ into a stream of pairs (shape
$[\omega, 2]$):
\begin{lstlisting}[mathescape, style=nonumbers]
imap [$\omega$]|[2] { _(iv): [a.[iv.[0],0], a.[iv.[0],1]]
\end{lstlisting}
Conceptually, an array of shape $[2, \omega]$ represents two infinite streams
that reside in the same data structure.  An operation on such a data structure
can progress independently on each stream, unless some dependencies on the outer
index are introduced.  A transpose, as above,
makes it possible to introduce such a dependency,
ensuring that the operations on both streams are synchronized.
A key to achieving this is the ability to re-enumerate infinite structures,
and ordinal-based infinite arrays make this possible.

\paragraph{Ackermann function}

The true power of multidimensional infinite arrays manifests itself in definitions
of non-primitive-recursive sequences as data. Consider the Ackermann function,
defined as a multi-dimensional stream:
\begin{lstlisting}[mathescape, style=nonumbers]
letrec a = imap [$\omega$, $\omega$]  {_(iv): letrec m = iv.[0] in
                                letrec n = iv.[1] in
                                if m = 0 then n + 1
                                else if m > 0 and n = 0 then a.[m-1, 1]
                                else a.[m-1, a.[m,n-1]] in a
\end{lstlisting}

Such a treatment of multi-dimensional infinite structures enables simple
transliteration of recursive relations \emph{as data}.
Achieving similar recursive definitions when using cons-lists is possible,
but they have a subtle difference. Consider a Haskell definition of
the Ackermann function in data:
\begin{lstlisting}[mathescape, style=nonumbers, language=Haskell]
a = [[ if m == 0 then n+1
       else if m > 0 then a !! (m-1) !! 1
       else a !! (m-1) !! (a !! m !! (n-1))
     | n <- [0..]]
    | m <- [0..]]
\end{lstlisting}
We use two $[0..]$ generators for explicit indexing, even though at
runtime, all necessary elements of the list will be present.  The lack of
explicit indexes forces one to use extra objects to encode the correct
dependencies, essentially implementing \imap{} in Haskell.  Conceptually,
these generators constitute two further locally recursive data structures.
Whether they can be always can be optimised away is not clear.
Avoiding these structures in an algorithmic specification
can be a major challenge.

\paragraph{Game of Life}

As a final example, consider Conway's Game of Life.
First we introduce a few generic helper functions:
\begin{lstlisting}[mathescape]
(or) $\equiv$ $\lambda$a.$\lambda$b.if a then a else b
(and) $\equiv$ $\lambda$a.$\lambda$b.if a then b else a
any $\equiv$ $\lambda$a.reduce or false a
gen $\equiv$ $\lambda$s.$\lambda$v.imap s {_(iv): v
$\nwarrow$ $\equiv$ $\lambda$v.$\lambda$a.imap |a| {_(iv): if any (iv$\dot{+}$v ${\dot{\texttt{>=}}}$ |a|) then 0 else a.(iv$\dot{+}$v)
$\searrow$ $\equiv$ $\lambda$v.$\lambda$a.imap |a| {_(iv): if any (iv $\dot{<}$ v) then 0 else a.(iv$\dot{-}$v)
\end{lstlisting}
\verb|or| and \verb|and| encode logical conjunction and disjunction, respectively.
\verb|any| folds an array of boolean expressions with the disjunction,
and \verb|gen| defines an array of shape $s$ whose
values are all identical to $v$.
More interesting are the functions $\nwarrow$ and $\searrow$.
Given a vector $v$ and an array $a$, they shift all elements of $a$ towards decreasing indices or
increasing indices by $v$ elements, respectively.
Missing elements are treated as the value $0$.

Now, we define a single step of the 2-dimensional Game of Life in APL style\footnote{See this video
by John Scholes for more details: \url{https://youtu.be/a9xAKttWgP4}}:
two-dimensional array $a$ by:
\begin{lstlisting}[mathescape]
gol_step $\equiv$ $\lambda$a.
    letrec F = [$\nwarrow$ [1,1], $\nwarrow$ [1,0], $\nwarrow$ [0,1], $\lambda$ x. $\nwarrow$ [1,0] ($\searrow$ [0,1] x),
                $\searrow$ [0,1], $\searrow$ [1,0], $\searrow$ [1,1], $\lambda$ x. $\searrow$ [1,0] ($\nwarrow$ [0,1] x)]
    in letrec
       c = (reduce ($\lambda$f.$\lambda$g.$\lambda$x.f x $\dot{+}$ g x) ($\lambda$x.gen |a| 0) F) a
    in
       imap |a| { _(iv): if (c.iv = 2 and a.iv = 1) or (c.iv = 3)
                         then 1
                         else 0
\end{lstlisting}
We assume an encoding of a live cell in $a$ to be $1$, and a dead cell to be $0$.
The array $F$ contains partial applications of the two shift functions to two-element vectors
so that shifts into all possible directions are present.
The actual counting of live cells is performed by a function which
folds $F$ with the function
$\lambda f.\lambda g.\lambda x.f\ x + g\ x$. This produces
$c$, an array of the same shape as $a$, holding the numbers of live cells
surrounding each position.
Defining the shift operations $\nwarrow$ and $\searrow$  to insert $0$ ensures that
all cells beyond the shape of $a$ are assumed to be dead.

The definition of the result array is, therefore, a straightforward \imap{}, implementing
the rules of birth, survival and death of the Game of Life.

The most interesting aspect of this algorithm is the fact that there is no
restriction on the shape
of $a$. In our transfinite setting, we can provide an array
of shape $[\omega,\omega]$. With no changes to source code, we can
deal with an infinitely large plane. An infinite $a$ requires a lazy implementation
as demanded by our semantics of \funclom{}, but a finite case
offers a strict implementation as a possible alternative.

\section{Transfinite Arrays \emph{vs.} Streams}

Streams have attracted a lot of attention due to the many algebraic properties they expose.
{}\cite{concrete-stream} provides a nice collection of examples,
many of which are based on the observation that streams form an applicative functor.
Transfinite arrays are applicative functors as well, not only for
arrays of shape [$\omega$], but also for any given shape \progvar{shp}.
With definitions:
\begin{lstlisting}[mathescape, style=nonumbers]
pure $\equiv$ $\lambda$x.imap shp {_(iv): x
($\diamond$) $\equiv$ $\lambda$a.$\lambda$b.imap shp {_(iv): a.iv b.iv
\end{lstlisting}
we obtain for arbitrary arrays $u$, $v$, $w$, and $x$ of shape \progvar{shp}:
\begin{mathpar}
    (\text{pure}\ \lambda x.x) \diamond u == u
    \and
    (\text{pure}\ (\lambda f.\lambda g.\lambda x. f\ (g\ x))) \diamond u \diamond v \diamond w == u \diamond (v \diamond w)
    \and
    (\text{pure}\ f) \diamond (\text{pure}\ x) == \text{pure}\ (f\ x)
    \and
    u \diamond (\text{pure}\ x) == (\text{pure}\ (\lambda f.f\ x)) \diamond u
\end{mathpar}

This shows that arbitrarily shaped arrays of finite size have this property,
as also shown by~\cite{Gibbons2017}, and that these properties can be expanded
into ordinal-shaped arrays.  Classical streams are a special instance of these,
\ie{} arrays of shape [$\omega$].

For stream operations that insert or delete elements, it is less obvious whether
these can be easily extended into ordinal-shaped arrays other than shape
[$\omega$].  As an example, let us consider the function \filter{}, which
takes a predicate $p$ and a vector $v$ and returns a vector that contains only
those elements $x$ of $v$ that satisfy $(p\ x)$.  A direct definition of
\filter{} can be given as:
\begin{lstlisting}[mathescape, style=nonumbers]
filter $\equiv$ $\lambda$p. $\lambda$v. if (p v.[0]) then v.[0] $\plusplus$ filter p (tail v)
                              else filter p (tail v)
\end{lstlisting}
This definition, in principle, is applicable to arrays of any ordinal shape, but
the use of \textit{tail} in the recursive calls inhibits application beyond
$\omega$.  Furthermore, the strict semantics of \funclom{} inhibits a
terminating application to any infinite array, including arrays of shape
$[\omega]$.  For the same reason, a definition of \filter{} through the
built-in \reduce{} is restricted to finite arrays.

To achieve possible termination of the above definition of \textit{filter}
for transfinite arrays, we would need to change 
to a lazy regime for all function applications
in \funclom{} and we would need to change the semantics of \imap{} into a variant
where the shape computation can be delayed as well.
Even if that would be done, we would still end up with an unsatisfying solution.
The filtering effect would always be restricted to the elements
before the first limit ordinal $\omega$.
This limitation breaks several fundamental properties, like those defined
in~\cite{theory-of-lists}, that hold in the finite and stream cases.
As an example, consider distributivity of filter over concatenation:
\begin{equation}\label{eqn:filtercon}
\filter\ p\ (a \plusplus b) == (\filter\ p\ a) \plusplus (\filter\ p\ b)
\end{equation}
This property holds for finite arrays, but fails with the
above definition of \filter{} in case $a$ is infinite.

To regain this property for transfinite arrays, we need to apply \filter{} to
all elements of the argument array, not only those before the first limit
ordinal $\omega$.  When doing this in the context of \funclom{}, the
necessity to have a strict shape for every object forces us to ``guess''
the shape of the filtered result in advance.  The way we ``guess'' has an impact
on the filter-based equalities that will hold universally.

In this paper we propose a scheme that respects the above
equality.  For finite arrays \filter{} works as usual, and for the
infinite ones, we postulate that the result of filtering will be of an
infinite-shape:
\[
    \forall p\forall a: |a| \geq \omega \implies |\filter\ p\ a| \geq \omega
\]
This is further applied to all infinite sequences contained within the given
shape as follows:
\[
    \forall i < | a |:
    ( \exists \lo\ \alpha : i < \alpha \leq |a| )
    \implies
    ( \exists k \in \mathbb{N} : p\ (a.(i+k)) = \true{} )
\]
We assume that each infinite sequence contains infinitely many elements for
which the predicate holds.  Consequently, any limit ordinal component of the
shape of the argument is carried over to the result shape and only any potential
finite rest undergoes potential shortening.
Consider a filter operation, applied to a vector of
shape $[\omega\cdot 2]$. Following the above rationale,
the shape of the result will be $[\omega\cdot 2]$ as well.  This means that the
result of applying \filter{} to such an expression
should allow indexing from $\{0, 1,\dots\}$ as well as from
$\{\omega,\omega+1,\dots\}$ delivering meaningful results.

This decision can lead to non-termination when there are only finitely many
elements in the filtered result.  For example:
\begin{lstlisting}[mathescape, style=nonumbers]
filter ($\lambda$x.x > 0) (imap [$\omega$+2] {_(iv): 0)
\end{lstlisting}
reduces to an array of shape $[\omega]$, which effectively is empty. Any
selection into it will lead to a non-terminating recursion.

The overall scheme may be counter-intuitive, but it states that for every index position
of the output, the computation of the corresponding value is well-defined.

Assuming the aforementioned behaviour of \filter{}, Eq.~\ref{eqn:filtercon}
holds for all transfinite arrays.
Another universal equation that holds for all transfinite vectors
concerns the interplay of \filter{} and \progvar{map}:
\begin{mathpar}
\filter\ p\ (\map\ f\ a) == \map\ f\ (\filter\ (p \cdot f)\ a)
\end{mathpar}

The proposed approach does not only respect the above equalities, but it also behaves
similarly to filtering of streams that can be found in languages such as
Haskell: \filter{} applied to an infinite stream cannot return a finite result.

\begin{figure}[h!]
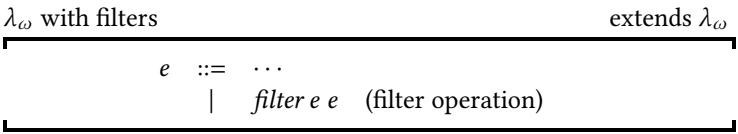

\begin{syntaxbox}[b]{.7\textwidth}%
                    {\funclom{} with filters\hfill extends \funclom{}}
\[
\begin{array}{ccll}
    e &::=& \cdots & \\
      &|& \filter\ e\ e & \text{(filter operation)}
\end{array}
\]
\end{syntaxbox}
\caption{\label{fig:syntaxomegafil}The syntax of \funclom{} with filters.}
\end{figure}

In principle, the chosen filtering scheme can be defined in \funclom{} by using
the \lo{} predicate within an \imap{}.  However,  the resulting definition is
neither concise, nor likely to be runtime efficient.  Given the importance of
\textit{filter}, we propose an extension of \funclom{}.
Fig.~\ref{fig:syntaxomegafil} shows the syntactical extension of \funclom{}.

As \filter{} conceptually is an alternative means of constructing
arrays, its semantics is similar to that of \imap{}. 
In particular, it constitutes a lazy array constructor, whose elements are being
evaluated upon demand created through selections.
Technically,
this means that we have to introduce a new value to keep \filter{}-closures,
a rule that builds such a closure from \filter{} expression, and we need to 
define the selection operation that forces evaluation within the filter closure.

We introduce as new value for \filter{}-closures:
\[
    \closure{\filter\ p_f\ p_e
             \ {\begin{cases}
                  \alpha_1 & v^1_r\ v^1_i\\
                  \dots\\
                  \alpha_n & v^n_r\ v^n_i
                \end{cases}}}
\]
which contains the
pointer to the filtering function $p_f$, the shape of the argument we are
filtering over ($p_e$) and the list of partitions that consist of a limit
ordinal, and a pair of partial result and natural number: $v_r$ and $v_i$
correspondingly.

On every selection at index $[\xi+n]$, where $\xi$ is a limit ordinal or
zero, and $n$ is a natural number, we find a $\xi$ partition within the filter
closure or add a new one if it is not there.  Every partition keeps a
vector with a partial result of filtering ($v_r$), and the index ($v_i$)
with the following property: the element in the array we are filtering
over at position $\xi+(v_i-1)$ is the last element in the $v_r$, given
that $v_r > 0$.  This means that if $n$ is within $v_r$, we return
$v_r.[n]$.  Otherwise, we extend $v_r$ until its length becomes $n+1$
using the following procedure: inspect the element in $p_e$ at the position
$\xi+v_i$ --- if the predicate function evaluates to \true{}, append
this element to $v_r$ and increase $v_i$ by one, otherwise, increase
$v_i$ by one.

A formal description of this procedure can be found
in~\cite[Sec.~2.1.4]{lamom-sem}.

\section{Towards an Implementation}\label{sect:impl}

We used the semantics of \funclom{} as a blueprint for the implementation
of an interpreter, called Heh 
%(available in the anonymous supplementary materials).
available at \url{https://github.com/ashinkarov/heh}).
The interpreter, which serves as a proof of concept,
performs a literal translation of the semantic rules provided in the paper
into Ocaml code. All examples provided in the paper
can be found in that repository, and run, correctly, in the interpreter.

The implementation parses the program, evaluates it and prints the result.
We represent the storage $S$ from our semantics by a hash table of pointer-value bindings.
Environments $\rho$ are implemented as lists of variable-pointer pairs.  Pointers
and variables are strings and values are of an algebraic data type.
In the proof-of-concept interpreter, we never actively discard pointers or
variables; however we do share pointers
and we update \imap{}/\filter{} closures in place, in the same way as it is done
in the formal semantics.

We represent ordinals in Cantor Normal Form.
The algorithms for implementing operations on ordinals are
based on~\cite{ordinal-theory}.
In the same paper, we also find an in-depth
study of the complexities of ordinal operations:
comparisons, additions and subtractions have complexities $O(n)$, where $n$ is the minimum
of the lengths of both arguments; multiplications have the complexity $O(n\cdot m)$, where
$m$ and $n$ are the lengths of the two argument representations.

% The Cantor Normal Form helps
% a lot, specifically with the $\omega^\omega$ restriction discussed in Section 3, as
% any ordinal can be see as a list of exponent-coefficient pairs with exponents
% being natural numbers.  In~\cite{ordinal-theory} the complexity of ordinal
% operations has been studied in depth.   Whether this can be
% further improved, for example by applying techniques similar to Fast Fourier
% Transformation, is an open question.

The interpreter makes it possible to run all the examples described in this
paper. Additionally, the interpreter provides means for experimentation through the
incorporation of variants in the semantics of \imap{}:
two interpreter flags enable users to (i) avoid the
memoization of array elements completely, and (ii) to apply the
strict \imap{}-semantics instead
of the lazy one whenever arrays are of finite shape.
The implementation comes with about 100 unit tests.

\subsection{Performance considerations}

Having an interpreter for \funclom{} available
allows experimentation with ordinal indexing and transfinite definitions.
However, one of our initial aims, to enable efficient runtime execution on
parallel systems, is not demonstrated by Heh.
In the remainder of this section, we discuss several performance considerations
that show how we envision efficient parallel executions of \funclom{} to be possible.

\paragraph{Strictness}
As mentioned in Section~2, the design of \funclal{} closely matches that of
\sac{} which has been shown to deliver high-performance execution
on a variety of parallel machine
architectures~\cite{WiesGrelHasl+JEI12,SinkSchoBern+CCPE13}.
Since \funclom{} is largely an extension of \funclal{} to support infinite arrays,
we expect that programs that refrain from using infinite arrays can be mapped in
\sac{} programs and, thus, benefit from the compiler tool chain\footnote{
    See \url{http://www.sac-home.org} for further details.}
for getting high-performance parallel execution.
A prerequisite for this is that switching from the
lazy variant of the \imap{}-construct as defined for \funclom{}, to the strict version
of \imap{} from \funclal{}, is valid, \ie{} the switch does not change the semantics of a
program under consideration.
A comparison of the corresponding two semantic definitions in Section~2 shows that
this is legitimate, if and only if (i) the shape of the array that is constructed is finite,
(ii) the array is not recursively defined, and (iii) all elements of the array are
being accessed.
Criterion (i) is trivial to decide. Criterion (ii), while being undecidable in general,
in practice, can be approximated in most cases straightforwardly.
The third criterion is more difficult to approximate by means of analyses.
We identify two possible alternatives to conservative approximation:
\begin{itemize}
    \item programmers could be allowed to explicitly annotate strictness of
        \imap{}-constructs or just individual partitions of them.
        While this seems very effective, in principle, it comes with some drawbacks as well:
        if a programmer annotates too little strictness, there might be a
        noticeable performance penalty and any wrong annotations could lead to non-termination.
    \item some form of dynamic switch between strict and lazy modes of
        \imap{} evaluation could be implemented, speculatively evaluating some arguments
        to some extent.
\end{itemize}
The ability to have mixed strict and lazy imap semantics in Heh facilitates experimentation
in this regard.

\paragraph{Strict Recursion}
Even if criterion (ii) from the previous paragraph is not given, as long as the other two
criteria hold, a strict evaluation
is possible but it can no longer be performed in a data-parallel style because
of dependencies between the elements.  Given that it is known in advance that
the entire subspace of the \imap{} needs to be evaluated, the order of traversal
of the elements can dramatically impact performance of such an evaluation.
If all the dependencies between the elements in a recursive \imap{} are linear
with respect to index, then such a recursive \imap{} can be presented
in the polyhedral model as a loop-nest.  This would give a rise to 
very powerful optimisations that are well understood within polyhedral
compilation frameworks.  The question whether infinite specifications can
be handled by the polyhedral model as efficiently as finite ones remains open,
offering perspective for future work.

\paragraph{Data structures}
The current semantics prescribe that, when evaluating selections into a lazy
\imap{}, the partition that contains the index is split into a single-element
partition and the remainder.  This means that, as the number of selections
into the \imap{} increases, the structure that stores partitions of the
\imap{} will have to deal with a large number of single-element arrays.
Partitions can be stored in a tree, providing $O(\log n)$ look-up; however
triggering a memory allocation per every scalar can be very inefficient.
An alternate approach would be to allocate larger chunks, each
of which would store a subregion of the index space of an \imap{}.
When doing so, we would need to establish a policy
on the size of chunks and chose a mechanism on how
to indicate evaluated elements in a chunk.
Another possibility would be to combine the chunking with some strictness
speculation, as explained in the previous paragraph.
We could trigger the evaluation of the entire chunk whenever any element of the
chunk is selected.

\paragraph{Memory management}
An efficient memory management model is not obvious.  In case of strict
arrays, reference counting is known to be an efficient solution~\cite{Cann89,GerlckSchoIJPP06}.
For lazy data structures, garbage collection is usually preferable.  Most likely,
the answer lies in a combination of those two techniques.

The \imap{} construct offers an opportunity for garbage collection
at the level of partitions.   Consider a lazy \imap{} of boolean values
with a partition that has a constant expression:
\begin{lstlisting}[mathescape, style=nonumbers]
imap [$\omega$] {..., l <= iv < u: false, ...
\end{lstlisting}
Assume further that neighbouring partitions evaluate to \emph{false}.  In this
case, we can merge the boundaries of partitions and instead of keeping values in
memory, the partition can be treated as a generator.  However, an efficient
implementation of such a technique is non-trivial.

\paragraph{Ordinals}

An efficient implementation of ordinals and their operations is also essential.
Here, we could make use of the fact that \funclom{} is limited to ordinals up to
$\omega^\omega$.  For further details see~\cite[Sec.~4]{lamom-sem}

\section{Related Work}

%\subsection{Extending indexing domain}

Several works propose to extend the index domain of arrays to increase
expressibility of a language.
%\paragraph{Cardinal route}
A straightforward way to do this is to stay within cardinal
numbers but add a notion of $\infty$, similarly to what we have proposed
in $\funclal^\infty$.  Similar approach is described
in~\cite{apl-infinity}; in J~\cite{J} infinity is supported as a value,
but infinite arrays are not allowed.  As we have seen, by doing so we
lose a number of array equalities.

%\paragraph{Theory of Arrays}
In~\cite[page 137]{more-arrayth} we read: \blockquote[]{%
    A restriction of indices to the finite ordinal numbers is a needless
    limitation that obscures the essential process of counting and indexing.}
We cannot agree more.  {}\cite{more-arrayth} describes an axiomatic array theory
that combines set theory and APL\@.  The theory is self contained and
gives rise to a number of array equalities.  However, the question on how this
theory can be implemented (if at all) is not discussed.

%{IBM's APL2 claims to implement More's array theory. Not sure if this
%what you're getting at. Dyalog APL should be close enough to APL2 to
%let you run experiments, should you care. Finite arrays only need apply.}
%
% This paper doesn't talk about implementation and most importantly
% implementation of ordinal-indexed arrays...

%\paragraph{Arrays indexed by reals}
In~\cite{Taylor:1982:IIA:390006.802264} the authors propose to extend the domain
of array indices with real numbers.  More specifically, a real-valued function
gives rise to an array in which valid indices are those that belong to the
domain of that function. The authors investigate expressibility of such arrays
and they identify classes of problems where this could be useful, but neither provide a full
theory nor discuss any implementation-related details.

Besides the related work that stems from APL and the plethora of array languages that
evolved from it, there is an even larger body of work that has its origins in
lists and streams.
One of the best-known fundamental works on the theory of lists using ordered pairs can be found
in~\cite[sec. 3]{mccarthy-lisp}, where a class of S-expressions is defined.
The concepts of \emph{nil} and \emph{cons} are introduced, as well as
\emph{car} and \emph{cdr}, for accessing the constituents of \emph{cons}.

%\begin{itemize}
%    \item Atomic symbols are S-expressions;
%    \item \emph{nil}, an atomic symbol used to terminate lists is S-expression;
%    \item if $e_1$ and $e_2$ are S-expressions, so is $(e_1, e_2)$.
%\end{itemize}
%After that, a number of functions to manipulate the list are defined.  Most
%importantly:
%\begin{description}
%    \item[\emph{cons} ($e_1$, $e_2$)] constructs a pair $(e_1, e_2)$;
%    \item[\emph{car} ($e$)] (also commonly referred as \emph{head} ($e$))
%        gets a first element of the pair;
%    \item[\emph{cdr} ($e$)] (also commonly referred as \emph{tail} ($e$))
%        gets the second element of the pair.
%\end{description}

The Theory of Lists~\cite{theory-of-lists} defines lists abstractly as linearly
ordered collections of data.  The empty list and operations like length of
the list, concatenation, filter, map and reduce are introduced axiomatically.
Lists are assumed to be finite.  The questions of representation of this
data structure in memory, or strictness of evaluation, are not discussed.

%\todo[inline]{Mention that it is Haskell and Category-theory oriented.}
Concrete Stream Calculus~\cite{concrete-stream} introduces streams as codata.
Streams are similar to McCarthy's definition of lists, in that
they have functions \emph{head} and \emph{tail},
but they lack \emph{nil}.  This requires streams to be
infinite structures only.  The calculus is presented within Haskell,
rendering all evaluation lazy.

Coinduction and codata are the usual way to introduce infinite data structures
in programming languages~\cite{JKS12b,KS16a}.  Key to the introduction of codata
typically is the use of coinductive semantics~\cite{leroy09coind}.
In our paper,
the use of ordinals keeps the semantics inductive and deals with
infinite objects by means of ordinals.  In~\cite{Turner1995}, the author investigates
a model of a total functional language, in which codata is used to define infinite
data objects.

Streams are also related to dataflow models, such as~\cite{KPN,Petri62,Estrin1963}.
The computation graphs in the latter can be seen as recursive expressions
on potentially infinite streams.  As demonstrated
in~\cite{beck_plaice_wadge_2015}, there is a demand to consider
multi-dimensional infinite streams that cache their parts for better efficiency.

%\subsection{Implementation Related}

Two array representations, called \emph{push arrays} and \emph{pull arrays},
are presented in~\cite{pushpull}.
Pull arrays are treated as objects that have a length and an
index-mapping function; push arrays are structures that keep
sequences of element-wise updates.  The \imap{} defined here can be considered an advanced
version of a pull array, with partitions and transfinite shape.
The availability of partitions
circumvents a number of inefficiencies, (\eg{} embedded conditionals) of
classical pull arrays; the ordinals, in the context of the \imap{}-construct,
 enable the expression of streaming algorithms.

The \#Id language, presented in~\cite{Heller_1989}, is similar to \funclom{};
It combines the idea of lazy data structures with an eager execution context.

In~\cite{productive-prog,Mogelberg:2014}, the authors propose a system that makes it
possible to reason whether a computation defined on an infinite stream is
productive\footnote{The computation will eventually produce the next item,
    \ie{} it is not stuck.} --- a question that can be transferred directly to \funclom{}.
Their technique lies in the introduction of a clock abstraction
which limits the number of operations that can be made before a value must be
returned.  This approach has some analogies with defining explicit ``windows''
on arrays, as for example proposed in~\cite{Hammes1999}, or guarantees that
programs run in constant space in~\cite{Lippmeier2016}.

One of the key features of the array language described in this paper is the
availability of strict shape for any expression of the language.  A similar
effect can be achieved by encoding shapes in types.  Specifically in the
dependently-typed system, such an approach can be very powerful.  The work on
container theory~\cite{ABBOTT20053} allows a very generic description of indexed objects
capturing ideas of shapes and indices in types.  A very similar idea in the
context of arrays is described in~\cite{Gibbons2017}.  The work on dependent
type systems for array languages include~\cite{Slepak:2014,TROJAHNER2009643,
Xi:1998}.  Finally, a way to extend a type theory to include the notion of
ordinals can be found in~\cite{hancock2000ordinals}.

\section{Conclusions and Future Work}

This paper proposes \emph{transfinite arrays} as a basis for 
a simple applied $\lambda$-calculus \funclom{}.
The distinctive feature of transfinite arrays is their ability to capture
arrays with infinitely many elements, while maintaining structure
within that infiniteness.
The number of axes is preserved, and individual axes can contain
infinitely many infinite subsequences of elements.
This capability extends, into the transfinite space, 
many structural properties that hold for finite arrays.

The embedding of transfinite arrays into \funclom{} 
allows for recursive array definitions, offering an opportunity
to transliterate typical list-based algorithms, including algorithms on infinite lists
for stream processing, into a generic array-based form. 
The paper presents several examples to this effect, and provides some 
efficiency considerations for them. It remains to be seen if these
considerations, in practice, enable a truly unified view
of arrays, lists, and streams.

The array-based setting of \funclom{} allows this recursive style of defining
infinite structures to be taken into a multi-dimensional context,
enabling elegant specification of inherently multi-dimensional problems 
on infinite arrays.
As an example, we present an implementation of Conway's Game of Life which,
despite looking very similar to a formulation for finite arrays, is defined for
positive infinities on both axes.
Within \funclom{}, accessing neighbouring elements along both axes can be specified without
requiring traversals of nested cons lists.

We also present an implementation for the Ackerman function, using a 
2-dimensional transfinite array, one axis per parameter. 
The resulting code adheres closely to the abstract declarative
formulation of the function, while also implicitly generating a 
basis for a memoising implementation of the algorithm.

An interesting aspect of transfinite arrays is that ordinal-based 
indexing opens up an avenue to express transfinite induction in data
in very much the same way as \emph{nil} and \emph{cons} are duals to 
the principle of mathematical induction.
This can not be done easily using \emph{cons} lists,
as there is no concept of limit ordinal in the list data structure. It may be
possible to encode this principle by means of nesting, but then one would need a
type system or some sort of annotations to distinguish lists of transfinite
length from nested lists.  The \imap{} construct from the proposed formalism can
be seen as an elegant solution to this.

The fact that \imap{} supports random access and is powerful enough to capture list and
stream expressions opens up an exciting perspective for the implementation
of \funclom{}. When arrays are finite, it is possible to reuse
one of the existing efficient array-based implementations.  When arrays are
infinite, we can use list or stream implementations to encode \funclom{}, but at
the same time the properties of the original \funclom{} programs open the door
to rich program analysis.  We believe that many functional languages
striving for performance could benefit from the proposed design, at least 
when finiteness of arrays can be determined by program analysis.

Although the concept of transfinite arrays offers many
new and interesting possibilities, we note several practical aspects that 
would benefit from further investigation.
It is not yet clear what are the most efficient implementations
for our proposed infinite structures. Choices of representation 
affect both memory management design and the guarantees that our
semantics can provide. A type system for the proposed formalism is far from
obvious, with the main question being the decidability of useful ordinal
properties in a type system.  The first-order theory of ordinal addition
is known to be decidable~\cite{Buchi1990}, but more complex ordinal theories
can quickly get undecidable~\cite{Choffrut2002}.  To our knowledge, there
is no type system that natively supports the notion of ordinals.

Furthermore, a number of extensions to the proposed formalism are possible.
For example, it would be very useful to support streams that can terminate.
Currently the only way one could model this
in the proposed formalism is by introducing a new ``end of stream'' value,
and defining an infinite stream, where from a certain index, all further
values will be ``end of stream''.
%  However, such an approach implies that
% operations like reduction or filter need to return finite results when
% applied to infinite arrays.  For classical streaming applications
% it would be useful to know the maximum number of elements one need
% to keep at any given time --- streaming windows.  Solving all those questions
% forms a vision for our future research.

%% Acknowledgments
% \begin{acks}                            %% acks environment is optional
%                                         %% contents suppressed with 'anonymous'
%   %% Commands \grantsponsor{<sponsorID>}{<name>}{<url>} and
%   %% \grantnum[<url>]{<sponsorID>}{<number>} should be used to
%   %% acknowledge financial support and will be used by metadata
%   %% extraction tools.
%   This material is based upon work supported by the
%   \grantsponsor{GS100000001}{National Science
%     Foundation}{http://dx.doi.org/10.13039/100000001} under Grant
%   No.~\grantnum{GS100000001}{nnnnnnn} and Grant
%   No.~\grantnum{GS100000001}{mmmmmmm}.  Any opinions, findings, and
%   conclusions or recommendations expressed in this material are those
%   of the author and do not necessarily reflect the views of the
%   National Science Foundation.
% \end{acks}

%% Bibliography
\bibliographystyle{plain}
\bibliography{ms}

%% Appendix
\appendix
%\input{proofs.tex}

%Text of appendix \ldots

\end{document}